\documentclass{article}

\usepackage{microtype}
\usepackage{graphicx}
\usepackage{subcaption}
\usepackage{booktabs} % for professional tables

\usepackage{hyperref}

\usepackage[accepted]{icml2026}

\usepackage{url}            % simple URL typesetting
\usepackage{amsfonts}       % blackboard math symbols
\usepackage{nicefrac}       % compact symbols for 1/2, etc.
\usepackage{xcolor}         % colors
\usepackage{amsmath,amsfonts,bm}

\def\eqref#1{equation~\ref{#1}}
\def\1{\bm{1}}

\DeclareMathAlphabet{\mathsfit}{\encodingdefault}{\sfdefault}{m}{sl}
\SetMathAlphabet{\mathsfit}{bold}{\encodingdefault}{\sfdefault}{bx}{n}

\usepackage{caption}
\usepackage[capitalize,noabbrev]{cleveref}
\usepackage{duckuments}
\usepackage{multirow}
\usepackage{makecell}
\usepackage{enumitem}
\usepackage{colortbl}
\usepackage{xspace}
\usepackage{lipsum}
\usepackage{amssymb}
\usepackage{mathtools}
\usepackage{amsthm}
\usepackage{bbm}
\usepackage{listings}
\usepackage{wrapfig}
\usepackage{float}
\usepackage{upquote} % replaced the unused code-listing package that required shell-escape (broke arXiv); upquote keeps verbatim quote glyphs identical
\usepackage{pifont}
\usepackage[most]{tcolorbox} 
\definecolor{promptgray}{gray}{0.95}
\usepackage[T1]{fontenc}

\Crefname{appsec}{Appendix}{Appendices}
\crefname{appsec}{Appendix}{Appendices}
\Crefname{appsubsec}{Appendix}{Appendices}
\crefname{appsubsec}{Appendix}{Appendices}
\Crefname{appsubsubsec}{Appendix}{Appendices}
\crefname{appsubsubsec}{Appendix}{Appendices}

\definecolor{RowHighlight}{gray}{0.9}

\theoremstyle{plain}

\theoremstyle{definition}

\theoremstyle{remark}

\usepackage[textsize=tiny]{todonotes}
\usepackage{duckuments}
\newcommand{\placeholder}[1]{{\color{lightgray}\lipsum[1]}}

\newcommand{\mname}{Action-Grounded Visual Memory (AGMem)\xspace}

\newcommand{\sname}{AGMem\xspace}

\usepackage[textsize=tiny]{todonotes}

\icmltitlerunning{Naive Visual Memory is Not Enough: A Failure-Mode Study of GUI Agents}

\begin{document}

\twocolumn[
  \icmltitle{
  % Failure Mode in Visual Memory for GUI Agents
  % Investigating the Effectiveness of Visual Memory for GUI Agents \\through the Lens of Failure Mode Analysis
  Naive Visual Memory is Not Enough: A Failure-Mode Study of GUI Agents
    }

  % It is OKAY to include author information, even for blind submissions: the
  % style file will automatically remove it for you unless you've provided
  % the [accepted] option to the icml2026 package.

  % List of affiliations: The first argument should be a (short) identifier you
  % will use later to specify author affiliations Academic affiliations
  % should list Department, University, City, Region, Country Industry
  % affiliations should list Company, City, Region, Country

  % You can specify symbols, otherwise they are numbered in order. Ideally, you
  % should not use this facility. Affiliations will be numbered in order of
  % appearance and this is the preferred way.
  \icmlsetsymbol{equal}{*}

  \begin{icmlauthorlist}
    \icmlauthor{Seoyoung Choi}{snu}
    \icmlauthor{Minseok Ko}{postech}
    \icmlauthor{Hyunseok Lee}{kaist}
    \icmlauthor{Kunwoong Kim}{kaist}
    \icmlauthor{Woomin Song}{kaist}
    \icmlauthor{Chanseok Jeon}{pion}
    \icmlauthor{Jinwoo Shin}{kaist}
  \end{icmlauthorlist}

  \icmlaffiliation{kaist}{KAIST}
  \icmlaffiliation{snu}{Seoul National University}
  \icmlaffiliation{postech}{POSTECH}
  \icmlaffiliation{pion}{Pion Corporation}

  \icmlcorrespondingauthor{Seoyoung Choi}{chlsy07@snu.ac.kr}
  \icmlcorrespondingauthor{Jinwoo Shin}{jinwoos@kaist.ac.kr}

  % You may provide any keywords that you find helpful for describing your
  % paper; these are used to populate the "keywords" metadata in the PDF but
  % will not be shown in the document
  \icmlkeywords{Machine Learning, ICML}

  \vskip 0.3in
]

% this must go after the closing bracket ] following \twocolumn[ ...

% This command actually creates the footnote in the first column listing the
% affiliations and the copyright notice. The command takes one argument, which
% is text to display at the start of the footnote. The \icmlEqualContribution
% command is standard text for equal contribution. Remove it (just {}) if you
% do not need this facility.

% Use ONE of the following lines. DO NOT remove the command.
% If you have no special notice, KEEP empty braces:
\printAffiliationsAndNotice{}  % no special notice (required even if empty)
% Or, if applicable, use the standard equal contribution text:
% \printAffiliationsAndNotice{\icmlEqualContribution}

\begin{abstract}

% Human interaction with Graphical User Interfaces (GUIs) is inherently processing to focus on local context:
% Human interaction with Graphical User Interfaces (GUIs) is inherently selective:
% users do not reason over every pixel of a screen, but abstract it into local visual states that determine what action should be taken next.
% % These states, such as icons, dialogs, selected objects, input fields, and error messages, also serve as cues for recalling how similar interactions were handled before.
% However, GUI agents built on Vision Language Models (VLMs) process screenshots as full images, lacking an explicit mechanism to focus on informative regions and recall from memory.
% Using such full-image memory further inflates the visual context and can introduce task-irrelevant visual evidence, degrading performance.
Graphical User Interface (GUI) agents are increasingly used to automate complex computer tasks across applications, websites, and operating systems. To improve their reliability, recent work has introduced experiential memory, where agents retrieve prior trajectories to guide decision-making in similar states. More recent approaches further extend this idea to visual memory by storing and retrieving screenshots from past interactions, providing agents with richer contextual information than text-only memories. However, the effect of visual memory in GUI agents remains insufficiently understood: it is unclear which failures visual memory mitigates, or which failures it exacerbates.
% To diagnose where this degradation originates, 
To systematically analyze the effect of visual memory, 
we introduce a taxonomy of four GUI agent failures (i.e., \emph{cognitive failure}, \emph{visual state misunderstanding}, \emph{hidden operation blindness}, and \emph{grounding error}) that map to distinct stages of the perception-reasoning-action pipeline.
We find that prepending full-image memory has a divergent effect on the failure distribution: it reduces state-level failures but worsens action-level ones, and increases hidden operation blindness and grounding error.
Motivated by this finding, we propose \mname, an action-grounded memory framework for GUI agents.
% We propose \mname, an action-grounded based memory framework for GUI agents.
The core idea of \sname is to store image crops that capture the local GUI region closely related to a successful action or a recovery, rather than storing full screenshots.
% To build such memory at scale, we develop an automated pipeline that extracts compact views from existing GUI trajectory datasets.
% At test time, the agent retrieves action-grounded memories conditioned on the current subtask and local observation, and uses them for planning, verification, and failure recovery.
Experiments on OSWorld show that \sname improves task success rates by 33.3 \% over full-image memory.
% while reducing visual context by 5.0\%p.
These results demonstrate that \sname is an effective
% and efficient 
representation for visual memory in GUI agents.

\end{abstract}
\section{Introduction}
\label{sec:intro}

% [GUI Agent]
Graphical User Interface (GUI) control is often determined by local visual states rather than by the entire screen.
These local states recur across tasks and applications, providing reusable cues for deciding what action should be taken next.
Recent advances in vision-language models (VLMs) have enabled GUI agents that operate directly from screenshots and natural-language instructions ~\citep{zhang2024large},
and prior work have scaled them by training on massive decision-making trajectories \citep{xu2024aguvis,wang2026opencua,wu2024atlas}
Despite this progress, modern GUI agents, including frontier models such as GPT~\citep{openai_gpt55_2026}, Claude~\citep{anthropic_opus47_2026}, and Gemini~\citep{google_gemini31pro_2026}, still fail in non-trivial ways, and the structure of these failures should receive attention.
% Prior GUI agents (e.g., OS-Atlas~\citep{wu2024atlas}, Aguvis~\citep{xu2024aguvis}, OpenCUA~\citep{wang2026opencua}) mostly focused on producing massive amount of decision trajectories and train the model with those datasets.
% While effective, these training-based approaches require substantial data collection and computation, and they cannot improve frontier models such as GPT~\citep{openai_gpt55_2026}, Claude~\citep{anthropic_opus47_2026}, and Gemini~\citep{google_gemini31pro_2026} whose weights are inaccessible.
% However, current VLM-based GUI agents typically consume screenshots as monolithic images, without an explicit representation of the local regions that determine actions.

% [related works and problem statement.]
\begin{figure*}[!t]
  \centering
  % (a) Cognitive Failure
  \begin{subfigure}[t]{0.45\textwidth}
    \centering
    \includegraphics[trim={0 0.7in 0 0}, clip, width=\textwidth]{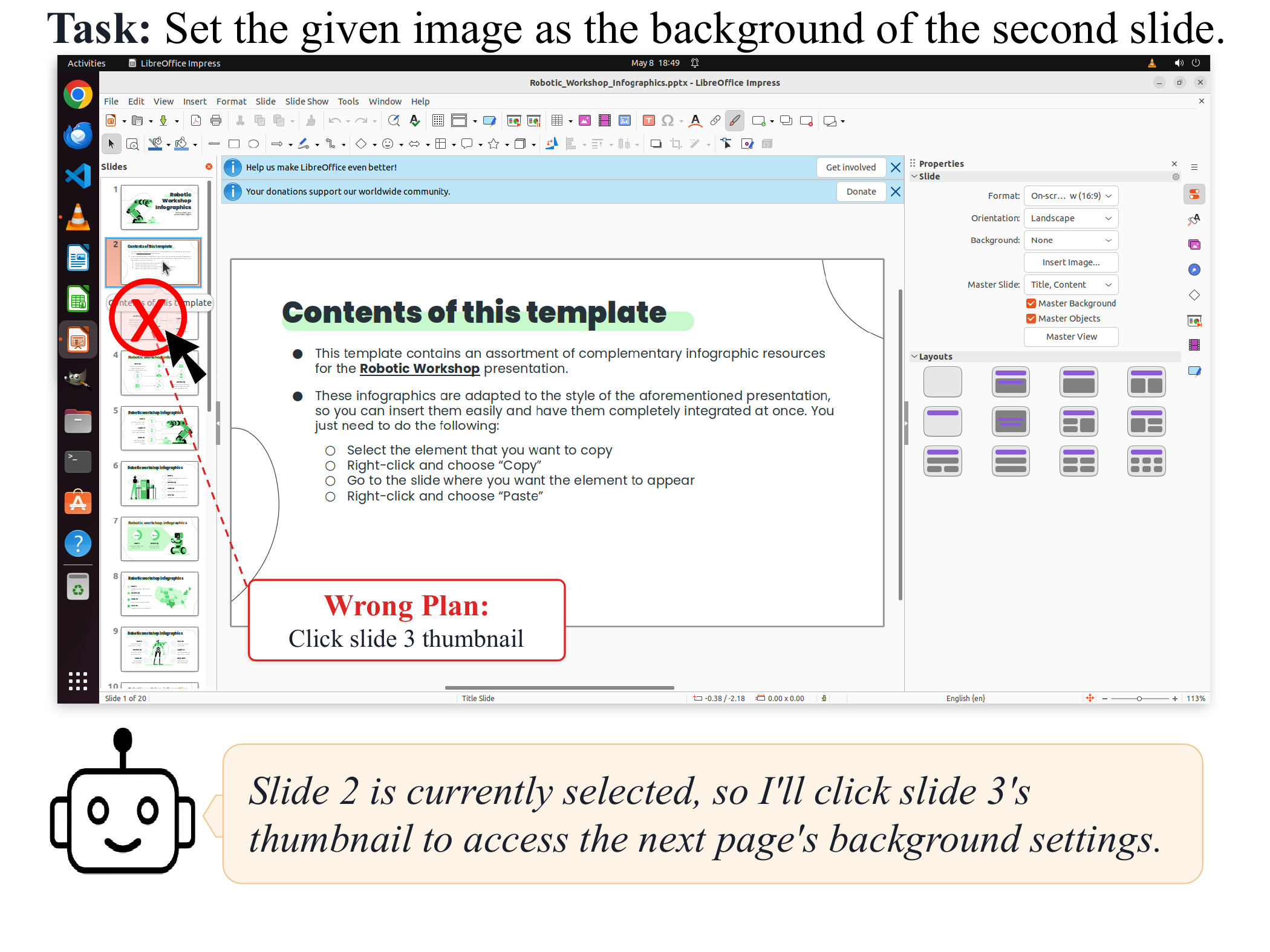}
    \caption{Cognitive Failure}
    \label{fig:teaser-a}
  \end{subfigure}
  \hspace{0.03\textwidth}
  % (b) Grounding Error
  \begin{subfigure}[t]{0.45\textwidth}
    \centering
    \includegraphics[trim={0 0.7in 0 0}, clip, width=\textwidth]{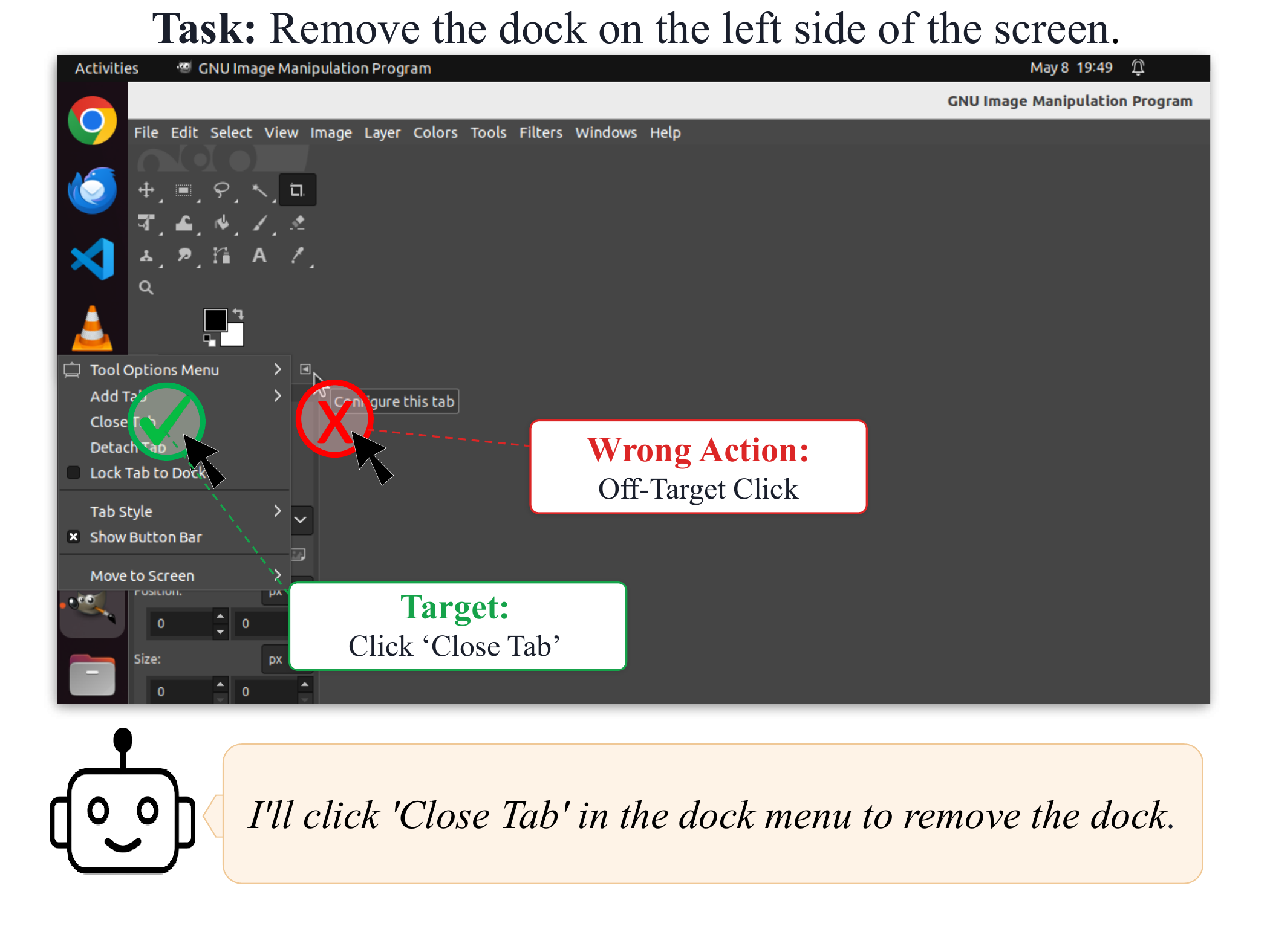}
    \caption{Grounding Error}
    \label{fig:teaser-b}
  \end{subfigure}
  
  \vspace{0.3em}
  
  % (c) Visual State Misunderstanding
  \begin{subfigure}[t]{0.45\textwidth}
    \centering
    \includegraphics[trim={0 0.7in 0 0}, clip, width=\textwidth]{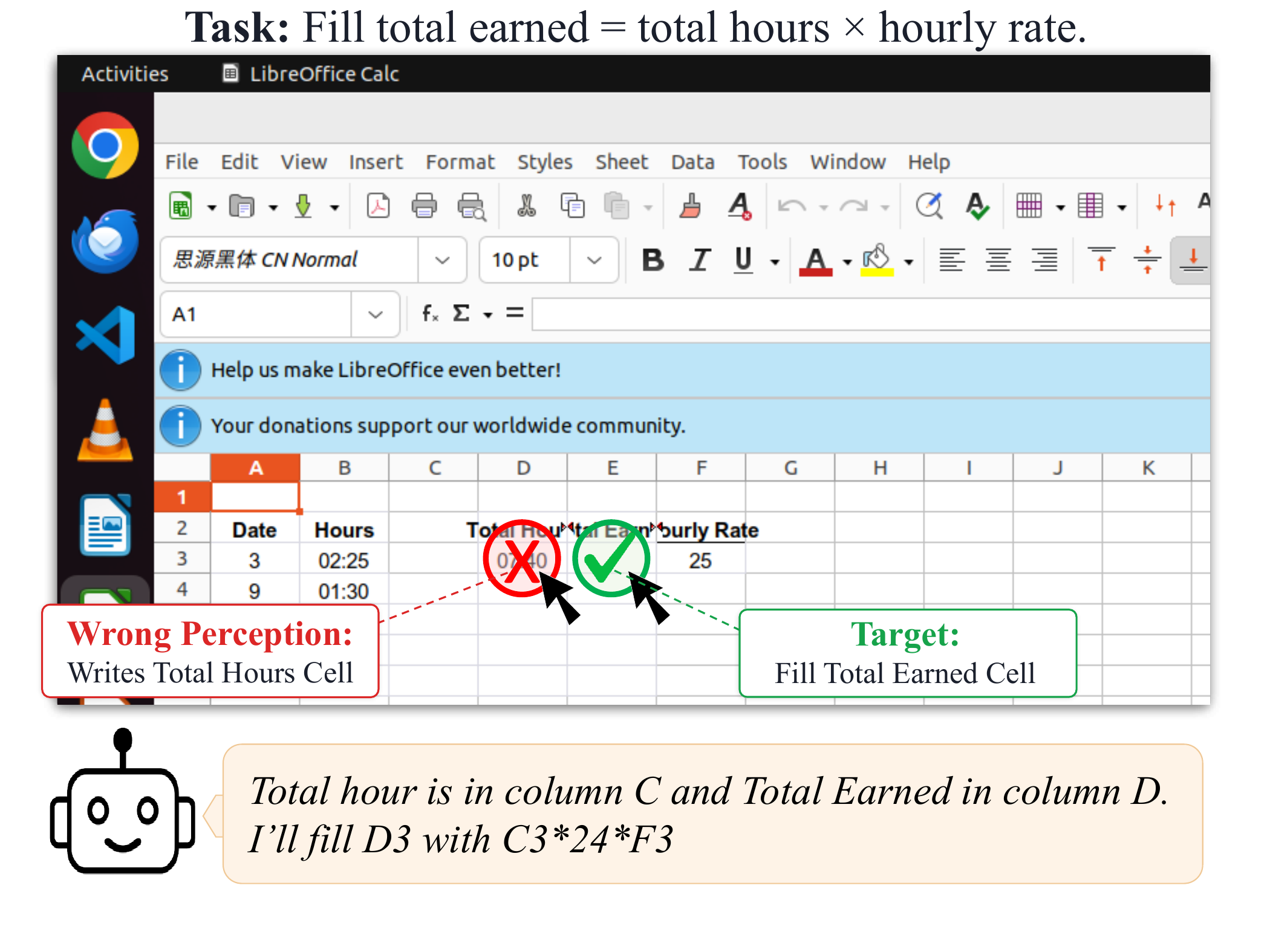}
    \caption{Visual State Misunderstanding}
    \label{fig:teaser-c}
  \end{subfigure}
  \hspace{0.03\textwidth}
  % (d) Hidden Operation Blindness
  \begin{subfigure}[t]{0.45\textwidth}
    \centering
    \includegraphics[trim={0 0.7in 0 0}, clip, width=\textwidth]{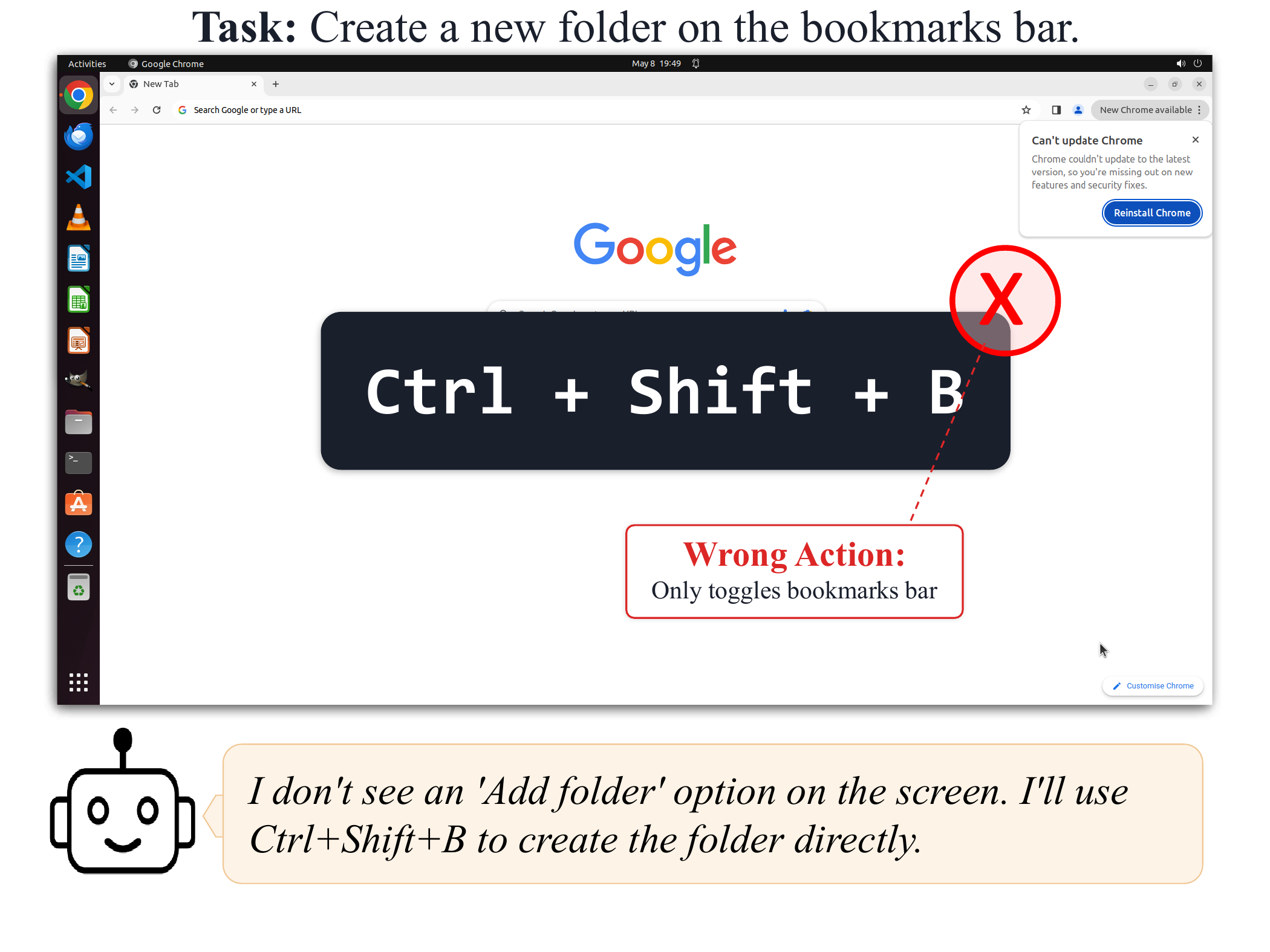}
    \caption{Hidden Operation Blindness}
    \label{fig:teaser-d}
  \end{subfigure}
  
  \caption{
    \textbf{Four failure modes of GUI agents.}
    We observe four recurring patterns in agent failures.
    In (\subref{fig:teaser-a}), the agent should be working on the second page, but instead clicks the thumbnail of the third page.
    In (\subref{fig:teaser-b}), the agent's target is correct, but the click is slightly off and misses the menu item.
    In (\subref{fig:teaser-c}), the agent misreads a cell, entering the formula into the Total Hours column instead of Total Earned.
    In (\subref{fig:teaser-d}), creating a folder requires an operation that is not visible on the screen, but the agent only presses \texttt{Ctrl+Shift+B}, a shortcut that toggles the bookmarks bar instead of creating the folder.
    Red and green markers denote the agent's action and the correct one, respectively.
  }
  \label{fig:teaser}
  \vspace{-0.05in}
\end{figure*}
% \input{resources/tables/error_analysis}  % moved to Sec.~\ref{sec:taxonomy}
% To tackle this problem, recent works proposed experiential working memory.
% Such methods log past decision trajectories and, at test time, retrieve the trajectories most relevant to the current state to guide action selection.
% One line of work is to summarize GUI screenshots, actions, and outcomes into textual memories~\citep{agashe2024agent, kagaya2024rap}.
% However, text-only memories discard information that is crucial for GUI control, such as icon appearance, layout, and multi-application environments.
% % While prior methods are promising, they overlook the most valuable information in GUI agents, namely, visualization, which can reduce their potential.
% Other works have attempted to develop image-based memory systems, e.g., \citet{wu2025auto}.
% % However, the mechanism how visual 
% However, they typically consume screenshots as monolithic images, without an explicit representation of the local regions that determine actions.
% % However, using full screenshots can substantially increase memory size and visual context length, which incurs intensive memory usage and cost inefficiency.
Recent works have proposed experiential working memory to tackle this problem. These methods log past decision trajectories and, at test time, retrieve the trajectories most relevant to the current state to guide action selection~\citep{agashe2024agent, kagaya2024rap}. More recent approaches further introduce image-based memory systems, such as \citet{wu2025auto}, which provide agents with richer, domain-specific contextual information beyond text. 
% In these systems, screenshots are commonly stored and retrieved as visual records of prior states. 
These systems typically consume screenshots as monolithic images, without an explicit representation of the local regions that determine actions.
Despite this progress, the role of visual memory in GUI agents remains insufficiently understood. Its effects have not been systematically characterized in terms of how it improves performance, which failure modes it mitigates, and where it may introduce regressions.

This leads to a question:
% \textit{How can GUI agents compactly represent visual memory and use them?}
\textit{What are the core failure modes in GUI agents, and how does visual memory mitigate them?}

% [Error Type categorization]
% \textbf{Our approach.} 
In this work, we categorize and analyze the common failure patterns of existing GUI agents.
Specifically, we investigate the failure patterns in representative benchmarks, including AgentNet \cite{wang2026opencua}, OSWorld \cite{OSWorld}, and WebForge~\cite{yuan2026webforge}.
On these benchmarks, we identify four recurring failure modes that together account for the majority of agent errors: (i) Visual State Misunderstanding, (ii) Hidden Operation Blindness, (iii) Cognitive Failure, and (iv) Grounding Error.

% \textbf{Taxonomy.} 
% First, \emph{Cognitive Failure} occurs when the agent misreads the task goal, current subtask, or success condition.
% Second, \emph{Visual State Misunderstanding} occurs when the agent misreads the currently visible screen state.
% Third, \emph{Hidden Operation Blindness} occurs when the correct action requires an operation not directly visible on the current screen.
% Lastly, \emph{Grounding Error} occurs when the agent intends the correct visible target but the executed coordinate or low-level action misses it.

% First, \emph{Cognitive Failure} occurs when the agent misunderstands the task goal, the current subtask, the required next step, or the success condition; for instance, the agent may follow the wrong plan step, assume a failed action succeeded, or terminate too early.
% Second, \emph{Visual State Misunderstanding} occurs when the agent misreads the currently visible screen state, such as believing a dialog or dropdown is open when it is not, or misidentifying the selected object.
% Third, \emph{Hidden Operation Blindness} occurs when the correct action requires an operation that is not directly visible on the current screen, such as opening a hidden menu, using a keyboard shortcut, or performing a non-obvious selection.
% Lastly, \emph{Grounding Error} occurs when the agent intends the correct visible target but the executed coordinate or low-level action misses it.

Based on these patterns, we then analyze how introducing visual memory shifts the failure distribution.
Following \citet{wu2025auto}, we prepend full screenshots of relevant history entries to the agent context.
For example, on OSWorld with GPT-5.4-mini, full-image memory reduces state-level failures (e.g., Visual State Misunderstanding 73.1\%$\to$69.6\%, but consistently worsens action-level failures, increasing Hidden Operation Blindness by 11.7\%p).
That is, full visual memory injects useful global context for state-level confusion but simultaneously crowds the visual context with task-irrelevant cues that distract from invisible operations and degrade coordinate grounding.
This suggests that visual memory is valuable in principle, but \emph{which pixels are stored} matters as much as whether memory is used at all.

To further mitigate this problem, we introduce \mname, an action-grounded information compacting framework for managing GUI agentic memory. The core idea of \sname is to enable agents to access focused GUI observations through a working memory and a domain-compacting retrieval system. Specifically, we introduce two focusing strategies:
(i) a visual memory consisting of cropped images that catch the elements that should be focused on, and
(ii) memory narrowing related to the given state.
To construct the memory system, we propose an automatic memory processing pipeline that can simply extract working memory trajectories from existing GUI action trajectories.
At test-time, the agent retrieves relevant crop-view memories and uses them during planning, allowing it to access decision-critical visual evidence while avoiding the cost of repeatedly processing full-screen images, as depicted in \cref{figure:method_overview}.
\begin{table*}[t]
\caption{\textbf{The four representative failure modes that can be observed in GUI agents.}}
\vspace{-0.04in}
\label{tab:failure_modes}
\centering
\small
\setlength{\tabcolsep}{6pt}
\renewcommand{\arraystretch}{1.15}
\begin{tabular}{p{0.18\linewidth}p{0.78\linewidth}}
\toprule
\textbf{Failure mode} & \textbf{Description} \\
\midrule
Cognitive failure & The model misunderstands the task goal, current subtask, required next step, success condition, or recovery strategy (e.g., following the wrong plan step, assuming a failed action succeeded, or terminating too early). \\
\addlinespace[2pt]
Visual state misunderstanding & The model misreads the visible screen state (e.g., believing a dialog or dropdown is open when it is not, misidentifying the selected object, or incorrectly inferring that a sort order). \\
\addlinespace[2pt]
Hidden operation blindness & The correct action requires an operation not directly visible until another UI action reveals it (e.g., opening a menu, context menu, or overflow menu; using a keyboard shortcut; performing a non-obvious selection). \\
\addlinespace[2pt]
Grounding error & The model intends the correct visible target and predicts a compatible action type, but the executed coordinate or low-level action misses the target. \\
\bottomrule
\end{tabular}
\vspace{-0.18in}
\end{table*}

% [Analysis of the effect of the new components]
In experiments, we demonstrate \sname consistently alleviates the full-image memory's failures in some taxonomies, (e.g., Visual State Misunderstanding reduces by 37.3\%p and Hidden Operation Blindness reduces by 26.3\%), while improving end-to-end task accuracy by 6.8\%p, on OSWorld with GPT-5.4-mini.
% Specifically, the fix improves the performance in xxx ways, but shows slight degradation in yyy aspects. This is because zzz. \placeholder{2}

% [Contribution summary]

We highlight the main contributions of this work as follows:
\begin{itemize}[itemsep=0.01in,leftmargin=0.2in,topsep=0.02in, nosep]
    \item{We introduce a four-class taxonomy of GUI agent failure modes and an LLM-as-Judge protocol for labeling them across online and offline benchmarks.}
    \item We provide the first empirical analysis of how visual memory shifts the failure distribution, showing that full-image memory cuts state-level failures but amplifies hidden-operation and grounding failures.
    % \item{We categorize the dominant failure modes in modern GUI agents.}
    % \item{We systemically analyze the empirical distribution of the failure modes, and investigate how introducing visual memory affects the distribution.}
    \item{We propose \sname, an action-grounded visual working memory framework for GUI agents with an automatic memory construction pipeline, which effectively mitigates the failure modes in GUI agents}
    % introduce action-grounded visual working memory framework for GUI agents, which preserves decision reasoning conditioned by visual states.}
    % \item{We propose an automatic memory construction pipeline that extracts action-grounded memory entries from existing GUI action trajectories.}
    % \item{We empirically show that \sname is useful ate mitigating failure mode xxx.}
\end{itemize}

\section{Failure modes of GUI Agents}
\label{sec:failure_modes}

In this section, we provide a comprehensive analysis of the trajectories of GUI agents and classify their failure modes.
First, we define a taxonomy of failure modes observed in the GUI agents (Section~\ref{sec:taxonomy}).
Then, we explain our setups (Section~\ref{sec:setup}).
Finally, we demonstrate main results of observed errors classified based on our taxonomy (Section~\ref{sec:main_results}).
% Finally, we provide detailed analysis on the failure modes (Section~\ref{sec:analysis}).

% To understand how GUI agents fail across environments, we conduct a benchmark-level failure-mode analysis on OSWorld~\citep{OSWorld}, AgentNetBench~\citep{wang2026opencua}, and WebForge~\citep{yuan2026webforge}. 
% Together, these benchmarks cover online/offline evaluation and desktop/web environments, which lets up seperate failure modes that are universal to GUI control from those tied to a particular interaction setting.

\subsection{Taxonomy: four failure modes}\label{sec:taxonomy}
We identify four recurring failure modes across benchmarks, as illustrated in \cref{fig:teaser}.
Each mode corresponds to a different stage of the perception-reasoning-action pipeline that a VLM-based GUI agent executes at every step: planning (\emph{cognitive failure}), perception (\emph{visual state misunderstanding}), action-space inference (\emph{hidden operation blindness}), and execution (\emph{grounding error}).
Because these stages require different kinds of evidence to fix (i.e., evidence about state for perception, about affordances for action-space inference, and about pixel-level alignment for execution), a memory mechanism that helps one stage need not help the other stages, which we exploit in the remainder of this section as a measurement instrument.

\begin{itemize}[itemsep=0.01in,leftmargin=0.2in,topsep=0.02in, nosep]
    \item \textbf{Cognitive failure.}
    This failure occurs when the model misunderstands the task goal, the current subtask, the required next step, the success condition, or the recovery strategy.
    It includes choosing the wrong action despite sufficient visible information, following the wrong plan step, assuming a failed action succeeded, terminating too early, or solving a semantically different task.

    \item \textbf{Visual state misunderstanding.}
    This is when the model misreads the current visible screen. 
    For example, the model may believe it is on the wrong page, think a dialog or dropdown is open when it is not, misidentify the selected object, or incorrectly infer that a form, filter, sort order, or progress state has already changed.

    \item \textbf{Hidden operation blindness.}
    The correct action requires an operation that is not directly visible until another UI action reveals it.
    Examples include opening a menu, toolbar, context menu, or overflow menu; using a shortcut; or performing a non-obvious selection or drag.

    \item \textbf{Grounding error.}
    This failure is observed when the model intends the correct visible target and predicts a compatible action type, but the executed coordinate or low-level action misses the target.
    We assign this type only when the raw reasoning indicates the same intended target as the ground truth; if the intended target is different, the failure is classified as cognitive failure or visual state misunderstanding instead.
\end{itemize}

\subsection{Setup}
\label{sec:setup}

\textbf{Evaluation setup.}
We evaluate GUI agents on three benchmarks:
two online benchmarks, OSWorld~\citep{xie2024osworld} and WebForge~\citep{yuan2026webforge}, and one offline benchmark AgentNetBench~\citep{wang2026opencua}.
Specifically, OSWorld contains 316 tasks across 10 application domains that use a $1920 \times 1080$ viewport. 
WebForge is evaluated on a randomly sampled subset of 50 tasks.
Each episode is limited to 50 steps unless explicitly stated.
Also, we evaluate AgentNetBench on pre-captured screenshot sequences and a given task instruction.
We report the average task score, computed from the benchmark’s composite action-matching metric aggregated over all tasks.

\textbf{Labeling setup.}
Based on our taxonomy in \Cref{tab:failure_modes}, we label failures with a Codex-based LLM-as-Judge that assigns a label only when the evidence is directly visible in the trace.
Note that the taxonomy is multi-label, so per-mode rates need not sum to $100\%$.
We label at the task level on OSWorld and WebForge, and at the action level on AgentNetBench.
The full details are given in Appendix~\ref{app:exp_details}.

\paragraph{GUI agent.}
A GUI agent is a large language model (LLM)-based autonomous agent that perceives and interacts with on-screen interface elements, such as screenshots, through human-like actions such as clicking and typing \cite{nguyen-etal-2025-gui}.
We follow the standard screenshot-based GUI agent setup~\citep{OSWorld, cheng2024seeclick}. Specifically, at each step $t$, an LLM-based policy receives the instruction $I$, a short history $h_t$ of recent screenshot-action pairs, the current screenshot $o_t$, and a set of retrieved memory steps $\mathcal{R}_t$ from a memory bank $\mathcal{M}$, and emits an action $a_t$:
$ a_t \sim \pi\bigl(\cdot \mid I,\, h_t,\, o_t,\, \mathcal{R}_t\bigr). $
% $$
%     a_t \sim \pi\bigl(\cdot \mid I,\, h_t,\, o_t,\, \mathcal{R}_t\bigr).
% $$
A full formalization of the policy and trajectory is deferred to \Cref{sec:appen-gui_intro}.

\begin{figure}[t]
\centering
\includegraphics[width=0.95\linewidth]{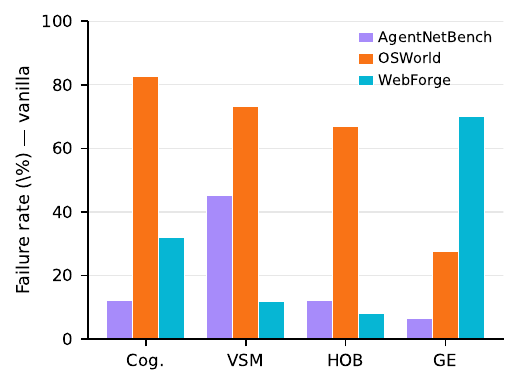}
\caption{\textbf{Failure mode distributions.} Failure mode distribution of GPT-5.4-mini across OSWorld, AgentNetBench, and WebForge.
Each benchmark exhibits a distinct dominant mode: hidden operation blindness on OSWorld, grounding error on WebForge, and visual state misunderstanding on AgentNetBench.}
\label{fig:cross_benchmark}
\vspace{-0.18in} 
\end{figure}

\subsection{Main results}
\label{sec:main_results}
\textbf{GUI agents frequently fall into four failure modes.}
Figure~\ref{fig:cross_benchmark} shows the vanilla failure distribution across the three benchmarks with GPT-5.4-mini.
Each environment has a distinct dominant mode: hidden operation blindness on OSWorld ($67.1\%$), grounding error on WebForge ($70.0\%$), and visual state misunderstanding on AgentNetBench ($45.1\%$).
Three modes-(i) cognitive failure, (ii) visual state misunderstanding, and (iii) grounding error-appear across all benchmarks, while hidden operation blindness is desktop-specific.
% This diversity sets up the question we ask in the rest of this section: does the effect of memory generalize across these very different failure profiles?

\textbf{Per-benchmark trends reflect environment characteristics.}
The dominant failure modes differ across benchmarks because each environment stresses a different part of the perception-reasoning-action pipeline.
OSWorld's high cognitive failure ($82.6\%$) and hidden operation blindness ($67.1\%$) reflect the long-horizon, menu-heavy nature of desktop applications, where essential commands are routinely buried in menus, context menus, ribbons, and keyboard shortcuts.
WebForge's grounding error dominance ($70.0\%$) reflects the dense layout of web interfaces with many small click targets, making pixel-level grounding more brittle than in larger desktop UIs; correspondingly, hidden operation blindness is rare ($8.0\%$) because web controls are typically exposed directly on the page.
AgentNetBench's lower absolute rates and the relative prominence of visual state misunderstanding ($45.1\%$) follow from its per-action evaluation, which isolates each step's state-recognition requirement and sidesteps long-horizon planning.
These various aspects of the benchmarks reflect the diverse modes of GUI agent usage and suggest where each evaluation is the most useful.
% These differences raise the question of whether the effect of memory generalizes across such different failure profiles, which we examine next.

\begin{figure}[t]
\centering
\includegraphics[width=0.95\linewidth]{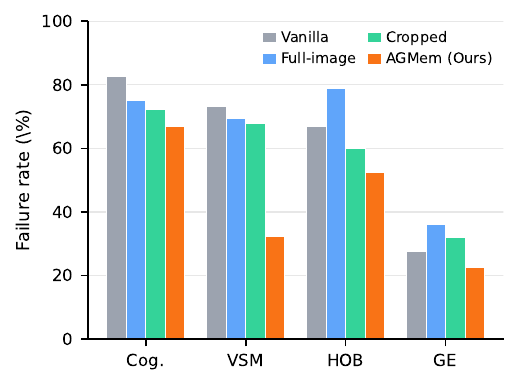}
\caption{\textbf{Per-mode failure rates on OSWorld with GPT-5.4-mini.} Full-image memory \emph{reduces} state-level failures (cognitive failure, visual state misunderstanding) but \emph{worsens} action-level failures (hidden operation blindness, grounding error).
% Cropping alone closes part of the gap.
\sname is the only configuration that consistently reduces all failure modes.}
\label{fig:failure_dist}
\vspace{-0.18in} 
\end{figure}
\begin{figure*}[t]
  \centering
  \includegraphics[trim={0in 0in 0in 0in}, width=0.8\textwidth]{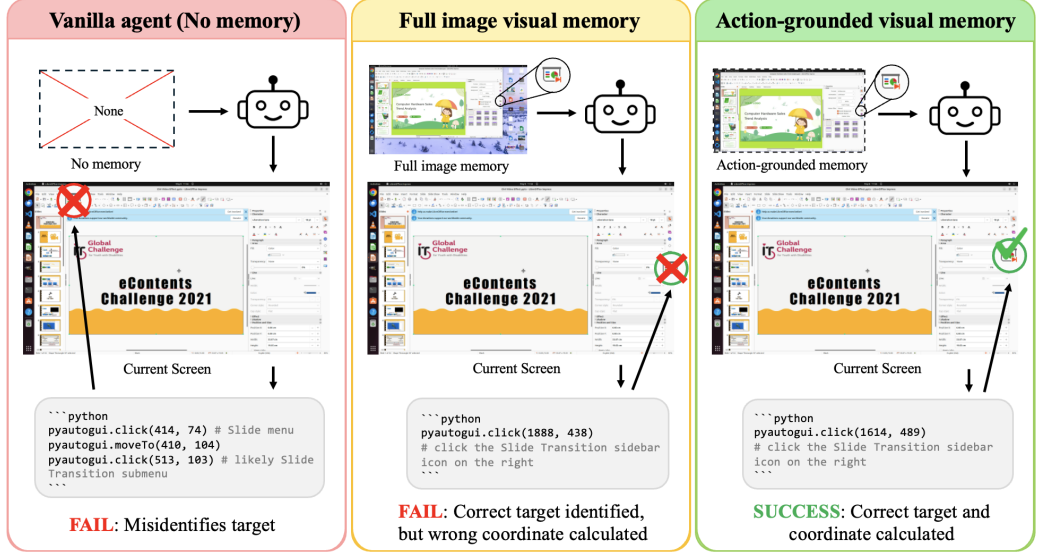}
  \caption{
  \textbf{Conceptual illustration of action-grounded visual memory.}
  Without memory, an agent may select an incorrect UI target.
  Full-image visual memory can provide useful prior experience, but the action-relevant cue may remain small or spatially ambiguous within the full screenshot, causing incorrect grounding.
  Action-grounded visual memory focuses the retrieved example on the region associated with the demonstrated action, making the relevant target easier to identify and localize.
  }
  \label{figure:failure_concept}
  % \vspace{-0.16in}
\end{figure*}

\textbf{Full-image memory has a divergent effect.}
As shown in Figure~\ref{fig:failure_dist}, adding full-image memory has a divergent effect on the failure distribution.
On the one hand, it reduces state-level failures by injecting global context that helps the agent re-orient to the task and the visible state: cognitive failure drops from $82.6\%$ to $75.0\%$ ($-7.6$\%p) and visual state misunderstanding drops from $73.1\%$ to $69.6\%$ ($-3.5$\%p).
On the other hand, the same full-screen retrieved screenshots substantially expand the visual context with task-irrelevant elements that distract the model from less salient operations.
This shows up as an increase in two action-level failures:
hidden operation blindness rises from $67.1\%$ to $78.8\%$ ($+11.7$\%p) and grounding error rises from $27.5\%$ to $36.1\%$ ($+8.6$\%p).
The net effect is that full-image memory trades state-level errors for action-level errors rather than consistent improvement.

\textbf{Directions for improvement.} This divergent effect points to a clear improvement target.
To preserve the gains in state-level failures while avoiding the costs on action-level failures, visual memory should be \emph{smaller} (containing only action-relevant pixels rather than full screenshots) and \emph{more selective} (narrowed to memory steps aligned with the current subtask).
We translate these two requirements into the design of \sname in the next section (Section~\ref{sec:method_analysis}).

\section{\sname: mitigating the side effects of visual memory} \label{sec:method_analysis}

% [TODO] Bridge from Sec.~\ref{sec:taxonomy}: motivate \sname components (action-grounded crops, recovery memory) as fixes for the dominant failure modes identified by the taxonomy.

\begin{figure*}[h!]
  \centering
  \includegraphics[width=0.8\textwidth]{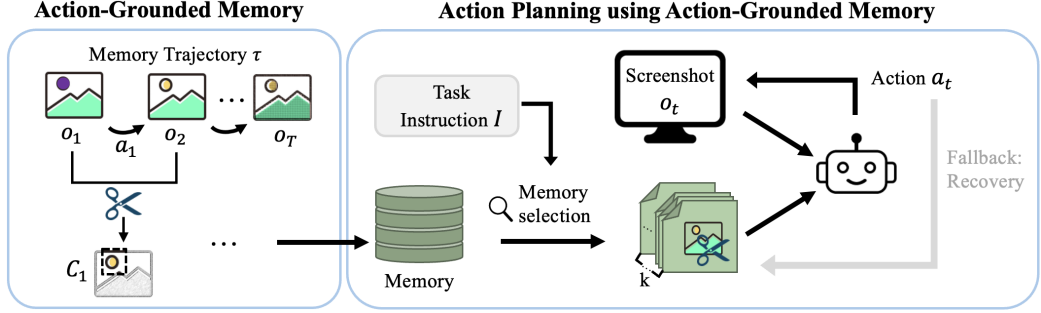}
  \caption{
  \textbf{Overview of \sname.}
  \sname is a compact GUI trajectory memory framework based on action-relevant visual views, salient memory candidates from trajectory- and step-level retrieval, and recovery memory for erroneous states.
  }
  \label{figure:method_overview}
\end{figure*}

% In this section, we present our proposed framework \mname, which addresses two directions for improvement as described in Section~\ref{sec:main_results}.
In this section, we present our proposed framework \mname, which addresses the two directions for improvement identified in Section~\ref{sec:main_results}.
First, we formalize the problem setup and outline \sname in \cref{subsec:overview}.
Then, we introduce action-grounded memory entries that compact each memory entry to an action-relevant view \Cref{subsec:cropping}.
Finally, we describe action-grounded memory candidates that narrow retrieval and provide a separate recovery memory for erroneous states \Cref{subsec:recovery}.
More details on \sname are provided in \Cref{app:algorithm_details}, including an overview in \cref{figure:method_overview}.

% In this section, we present our proposed framework \mname, a GUI agentic system that keeps GUI trajectory memory compact by retaining only salient parts.
% We first formalize the problem setup and briefly outline \sname in \cref{subsec:overview}.
% We then describe each component of \sname with details, in \cref{subsec:cropping,subsec:recovery}.
% \Cref{subsec:cropping} introduces \textit{salient memory entries}: each memory entry is reduced to an action-relevant view that retains only the screen region in which an action takes effect.
% \Cref{subsec:recovery} introduces \textit{salient memory candidates}: retrieval is narrowed to a small set of trajectories and steps aligned with the subtask structure, and a separate \textit{recovery memory} is used for recovery from erroneous states.
% An overview of \sname is depicted in \cref{figure:method_overview}.

%%%%%%%%%%%
\subsection{Overview}\label{subsec:overview}
To guide the agent through a given long-horizon instruction $I$, \sname first decomposes $I$ into a list of \textit{subtasks} \cite{gao2024assistguitaskorienteddesktopgraphical,ye2025mobileagentv3fundamentalagentsgui}:
$$ \mathbf{s} = ( s_1, s_2, \dots, s_p ). $$
Here, each $s_i$ is a natural-language subtask derived from the instruction $I$ alone before any action is executed.
Note that the number of subtasks $p$ is generally different from the number of steps $T$, since one subtask may span multiple steps.
The pair $(I, \mathbf{s})$ is given to the agent throughout the trajectory.
Before emitting each action, the agent self-reports its current subtask, so that action execution is aligned with the current subtask.

We then keep the memory salient in two core components.
First, an action-relevant view (\Cref{subsec:cropping}).
Second, a narrowed retrieval pipeline with a separate recovery memory for erroneous states (\Cref{subsec:recovery}).
 
% First (\cref{subsec:cropping}), every memory step is reduced to an \textit{action-relevant view}, a tight crop of the screenshot closely related to the executed action.
% This action-relevant view is used for every step stored in memory, so that retrieval and action prediction can focus on action-relevant views rather than full screenshots.
% Second (\cref{subsec:recovery}), retrieval is restricted to salient candidates in two stages.
% A trajectory-level retriever first selects a small pool of memory trajectories that cover the subtasks in $\mathbf{s}$.
% Then, a step-level retriever searches for the few memory steps closest to the present state.
% A recovery-state detector additionally redirects retrieval to a separate \textit{recovery memory} once the agent enters an erroneous state, providing examples of how to recover rather than only examples of successful execution.

%%%%%%%%
\subsection{Action-grounded visual memory}\label{subsec:cropping}

A desktop screenshot is often dominated by content unrelated to the current action or state (e.g., background windows, system trays, and idle panels), that is not affected by the agent's actions \cite{Lin_2025_CVPR, li2025screenspotpro}.
Storing all such screenshots would unnecessarily expand the memory bank and burden the agent's visual context with task-irrelevant cues.
We therefore reduce each memory entry to its salient form, called the \textit{action-relevant view}.
For each step $t$, we denote this view by $C_t$, which is a tightly cropped image in which action $a_t$ has important effects.

\paragraph{Construction from trajectories.}
Starting from screenshot $o_t$ at time $t$ (\cref{fig:crop_input}), the agent executes action $a_t$, and the environment transitions to the next screenshot $o_{t+1}$.
Given the consecutive screenshots $(o_t,o_{t+1})$, we construct the action-relevant view $C_t$, which is a tightly cropped snapshot that isolates the GUI region in which action $a_t$ takes effect. (\cref{fig:crop_output}).
See Appendix~\ref{app:algorithm_details} for the full algorithm.
% The full algorithm for construction process is given in Appendix~\ref{app:algorithm_details}.

% $$
%     M_t = \mathrm{ChangeMask}(o_t, o_{t+1}),
%     \qquad
%     C_t = \mathrm{Crop}\bigl(o_{t+1};\, \mathrm{Snap}(\mathrm{BBox}(M_t))\bigr).
% $$

% Given the consecutive screenshots $(o_t,o_{t+1})$, we construct the action-relevant view $C_t$ closely related to $a_t$ 

% This construction proceeds in four steps.
% First, we compute a pixel-wise difference between the two screenshots, subtract a locally smoothed version to suppress slowly varying background, and binarize the residual under a soft and a sharp threshold.
% We apply standard morphological opening and closing to remove noise and connect fragmented regions into a single change mask~\citep{bhutada2022opening} .
% Second, we take the tightest bounding box enclosing the change mask.
% Third, we snap the bounding box to the smallest UI container, such as a window, panel, or dialog returned by a UI parser, that fully contains it.
% Finally, if no qualifying container is returned by the parser, we keep the bounding box itself; if no change is detected at all, we fall back to the full screenshot.

% At test time, we use the previous step's view $C_{t-1}$ as the visual query for memory retrieval, which computed from the most recent pair $(o_{t-1}, o_t)$.
% (see \cref{subsec:recovery}).
% For $t = 0$, where no previous transition exists, we use the initial full screenshot $o_0$ directly as the query.

\paragraph{Memory structure.}
We construct the memory bank $\mathcal{M}$ from AgentNet dataset \cite{wang2026opencua}, $\mathcal{D}_{\text{AGNet}}$, where we start from retaining only non-redundant and correct trajectories $\tau$. 
% whose steps are annotated for redundancy and correctness.
% We retain only non-redundant and correct trajectories $\tau$ for the memory bank.

For each trajectory $\tau \in \mathcal{D}_{\text{AGNet}}$, an LLM produces a list of post-hoc subtask labels $\mathbf{s}_\tau$ that contains information about 
what actually happened on the screen during $\tau$, and we segment $\tau$ along $\mathbf{s}_\tau$ into sub-trajectories.
% we provide both the original instruction $I$ and the full trajectory $\tau$ to an LLM, which annotates $\tau$ with a list of \textit{subtask labels}:
% $$
%     \mathbf{s}_\tau = ( s_1, s_2, \dots, s_q ).
% $$
Consequently, the atomic unit of memory is a \textit{memory step}:
$$m_t = (s_j, a_t, C_t).$$
where $s_j\in\mathbf{s}_\tau$ is the subtask label that the step realizes, $a_t$ is the executed action, and $C_t$ is the action-relevant view.
Sub-trajectory segmentation and indexing details are given in Appendix~\ref{app:algorithm_details}.

% In contrast to $\mathbf{s}$, which is derived from the instruction $I$ alone before any action is executed, $\mathbf{s}_\tau$ is generated post-hoc, after $\tau$ has been observed in full.
% Hence, it is also grounded in the executed trajectory $\tau$ as well as the instruction $I.$
% Thus, its segmentation reflects what actually happened on the screen through the trajectory $\tau$, rather than only what the user asked for through $I$.

% We then split $\tau$ along $\mathbf{s}_\tau$ into \textit{sub-trajectories}:
% $$
%     \tau = \tau_{s_1} \oplus \tau_{s_2} \oplus \cdots \oplus \tau_{s_q},
% $$
% where $\tau_{s_j}$ is the segment of screenshot-action pairs realizing $s_j \in \mathbf{s}_\tau$ and $\oplus$ denotes sequence concatenation.

% We further annotate every step of every trajectory with its action-relevant view $C_t$.
% The atomic unit of memory is therefore a \textit{memory step}: a triple consisting of
% (i) the subtask label $s_j \in \mathbf{s}_\tau$ that the step realizes,
% (ii) the executed action $a_t$, and
% (iii) the action-relevant view $C_t$.
% Thus, $\mathcal{M}$ stores trajectories with subtask labels, sub-trajectories, and per-step memories.

%%%%%%%%
\subsection{
% Salient memory candidates: retrieval and recovery
Recovery-aware memory retrieval
}\label{subsec:recovery}

Searching the entire memory bank at every step or action is both costly and noisy, since most trajectories are irrelevant to the current task and individual steps may produce spurious visual memories.
We therefore narrow the search space in two stages.
We first select a small pool of trajectories at the task level, and then retrieve individual memory steps within that pool.
In addition, when the agent flags an error, we redirect retrieval to a separate \textit{recovery memory}.

\paragraph{Two-stage retrieval}
Given the current subtask list $\mathbf{s}$, a trajectory-level retriever first selects a size-$k$ pool $\mathcal{M}_{\text{sub}} \subseteq \mathcal{M}$ of memory trajectories whose post-hoc subtask labels $\mathbf{s}_\tau$ are close to $\mathbf{s}$ in a Sentence-Transformer~\citep{reimers-2019-sentence-bert} embedding space, greedily covering each subtask in $\mathbf{s}$.
A step-level retriever then encodes each memory step's subtask label and action-relevant view with the CLIP encoders~\citep{pmlr-v139-radford21a},
% ($e_T$ for text, $e_I$ for image),
computes a fixed-weight similarity against the current state -- the agent's self-reported subtask $s_t$ and the previous step's view $C_{t-1}$ -- and returns the top-5 memory steps as $\mathcal{R}_t$.
Encoder choices and more details are given in \Cref{app:algorithm_details}.

\paragraph{Recovery-aware memory}
A common reason behind the failures of GUI agents is error propagation, where the agent enters an erroneous state and the subsequent action proposals are made under a faulty premise.
To address this issue, we add recovery-aware verification memory which teaches the agent \emph{how to recover}.
Specifically, LLM-based \textit{recovery-state detector} detects the erroneous step during operation.
When an error is detected, the recovery memory provides examples of correct behaviors that bring the agent back from a faulty state to a valid one.
The recovery memory is built from AgentNet sub-trajectories that were excluded from the main bank (redundant or incorrect) but were later corrected within the same trajectory.
Specifically, each entry pairs a faulty sub-trajectory with the correct step, labeled by failure mode so retrieval can match the flagged pattern.
Construction details are given in \Cref{app:algorithm_details}.

% One of the common failure modes of GUI agents is the error propagation.
% Once the agent enters an erroneous state, its subsequent action proposals are made under a faulty premise.
% This issue is difficult to handle with the main memory bank alone, because the main memory bank only stores successful executions and therefore does not contain examples of `how to recover' from mistakes.

% To address this issue, we propose to use an LLM-based \textit{recovery-state detector}.
% Before each proposed action is executed, the detector inspects the recent screenshots, the executed actions, and the proposed action.
% It then predicts whether the agent is in an erroneous state and, if so, identifies the corresponding failure type (\cref{tab:failure_classification} in \cref{app:recovery_memory}).
% When no error is detected, the proposed action is executed as usual.
% When an error is detected, the proposed action is blocked, and retrieval is redirected to a separate \textit{recovery memory}.
% This recovery memory provides examples of corrective behaviors that bring the agent back from a faulty state to a valid one.

% The recovery memory is constructed from AgentNet sub-trajectories that were excluded from the main memory bank as redundant or incorrect, but were later corrected within the same trajectory.
% Each recovery example makes a pair of a faulty sub-trajectory and the corrective step that restores a valid state.
% Each example is also labeled by failure type, allowing retrieval to match the flagged failure pattern rather than rely only on global similarity.

\section{\sname experiments}\label{sec:experiments}

We provide empirical evidence of \sname by analyzing how it improves the failure mode distribution that full-image memory induces (Section~\ref{sec:agm_failuremode}), and end-to-end task performance against existing baselines 
(Section~\ref{sec:agm_main}).

\begin{table*}[t]
  \centering\small
  \caption{\textbf{Failure Mode Analysis on OSWorld Benchmark.} 
  Comparison of overall Success Rate (Acc.) and the distribution of specific failure modes across different visual memory configurations.}
  \label{tab:failure_analysis}
  \vskip 0.1in
  \footnotesize
  \setlength{\tabcolsep}{10pt}
  % \resizebox{0.65\linewidth}{!}{
  \begin{tabular}{l c cccc}
    \toprule
    \multirow{2}{*}{\textbf{Method}} & \textbf{Overall} & \multicolumn{4}{c}{\textbf{Failure Mode (Root Cause \%)}} \\
    \cmidrule(lr){3-6}
    & \textbf{Acc. (\%)} & \textit{Cognitive} & \textit{Visual Mis.} & \textit{Hidden Op.} & \textit{Grounding} \\
    \midrule
    Vanilla Agent & 18.3 & 82.6 & 73.1 & 67.1 & 27.5 \\
    \phantom{00}+ Visual Memory & 20.4 & 75.0 & 69.6 & 78.8 & 36.1 \\
    \phantom{0000}+ Crop & 25.8 & 72.4 & 68.3 & 60.3 & 32.1 \\
    \rowcolor{RowHighlight}
    \textbf{AGMem (Ours)} & \textbf{27.2} & \textbf{67.1} & \textbf{32.3} & \textbf{52.5} & \textbf{22.5} \\
    \bottomrule
  \end{tabular}
  % }
  \vspace{0.05in}
  \begin{flushleft}
    \scriptsize * \textit{Cognitive}: Cognitive Failure / \textit{Visual Mis.}: Visual State Misunderstanding / \textit{Hidden Op.}: Hidden Operation Blindness / \textit{Grounding}: Grounding Error
  \end{flushleft}
  \vspace{-0.10in}
\end{table*}
\subsection{Failure mode analysis}\label{sec:agm_failuremode}
We use the failure mode taxonomy of Section~\ref{sec:taxonomy} to measure how \sname is behaving on the failure distribution, on OSWorld with GPT-5.4-mini.

\textbf{\sname reduces every failure modes.}
As shown in Table~\ref{tab:failure_analysis}, \sname consistently reduces all four failure modes against the vanilla agent. For example, visual state misunderstanding reduces by about -40.8\%p.
This suggests that compact action-relevant crops are particularly effective at processing the small subset of pixels that determine state-level decisions.
Notably, while hidden operation blindness and grounding error are the two failure modes that are worsened by full-image memory, \sname also reduces them.
This indicates the gains from memory persist even when the visual context is restricted to action-relevant regions.

\textbf{Cropping alone is insufficient.}
The cropped configuration crops memory screenshots but performs retrieval over the full memory bank.
Compared to full-image memory, cropping alone partially recovers performance on hidden operation blindness ($78.8\%\to60.3\%$) and grounding error ($36.1\%\to32.1\%$), but still falls short of \sname in every mode and especially on visual state misunderstanding ($68.3\%$ vs.\ $32.3\%$).
The remaining gap is closed by subtask-aligned retrieval narrowing, which restricts retrieval to memory steps relevant to the current subtask and prevents off-task crops from re-injecting irrelevant cues.

% \placeholder

% \textbf{The pattern generalizes across benchmarks.}
% \placeholder

\newcolumntype{C}[1]{>{\centering\arraybackslash}p{#1}}
\begin{table}[t]
  \centering\small
  \caption{\textbf{\sname improves performance by effectively leveraging visual memory.}
  Accuracy (\%) and Milestone Accuracy (M. Acc.) of \sname (Ours) compared with baselines across three benchmarks, for GPT-5.4-mini. 
  For OSWorld and WebForge, we report overall success rate; for other, step-level accuracy.}
  % \label{tab:merged_results}
  \label{tab:main_results}
  \vskip 0.1in
  \footnotesize 
  \setlength{\tabcolsep}{8pt} 
  % \adjustbox{width=\linewidth}{
  \resizebox{\linewidth}{!}{
  \begin{tabular}{l c c cc}
    \toprule
    \multirow{2}{*}{\textbf{Method}} 
    & \textbf{OSWorld} 
    & \textbf{WebForge} 
    & \multicolumn{2}{c}{\textbf{AgentNet}} \\
    \cmidrule(lr){2-2} \cmidrule(lr){3-3} \cmidrule(lr){4-5}
    & Acc. & Acc. & Acc. & M. Acc. \\
    \midrule

    % \textit{GPT-5.4-mini}\vspace{0.02in} & & & & \\
    Vanilla agent 
    & 18.3 & 2.0 & 25.8 & 24.2 \\
    \phantom{0}+ visual memory 
    & 20.4 & 2.0 & 28.1 & 33.8 \\
    \rowcolor{RowHighlight}
    \textbf{\sname (Ours)} 
    & \textbf{27.2} & \textbf{2.0} & \textbf{28.8} & \textbf{34.6} \\
    % \midrule

    % \textit{Qwen3.6-35B-A3B}\vspace{0.02in} & & & & \\
    % \phantom{00}Vanilla agent 
    % & \textbf{36.0} & 36.0 & 59.6 & 66.6 \\
    % \phantom{0000}+ visual memory 
    % & 22.0 & 36.0 & 64.1 & 70.7 \\
    % \rowcolor{RowHighlight}
    % \phantom{00}\textbf{\sname (Ours)} 
    % & 34.0 & \textbf{42.0} & \textbf{66.2} & \textbf{71.6} \\
    \bottomrule
  \end{tabular}
  }
\end{table}
\subsection{Main results} \label{sec:agm_main}

We first present the performance of \sname by comparing GUI agent performance with other baselines. Here we mainly compare with one offline benchmarks (i.e., AgentNetBench) and two online benchmarks (i.e., OSWorld and WebForge). Furthermore, we demonstrate the token efficiency of \sname by comparing average token consumption per sample with the visual memory baseline against other baselines.

\textbf{Significant performance improvement.} As shown in Table~\ref{tab:main_results}, \sname consistently outperforms the existing baselines across both offline and online benchmarks, with large margins.
For instance, on OSWorld, \sname improves task accuracy by +8.9\%p (from 18.3\% to 27.2\%) with GPT-5.4-mini.
On AgentNet Bench, \sname further improves step accuracy from 25.8\% to 28.8\% with GPT-5.4-mini.
These results show that \sname is not specialized to a single benchmark or interaction setting, but generalizes across both web-based GUI tasks and online desktop-control environments.
This suggests that compact and task-relevant visual memory can provide useful state information for GUI agents without relying on full visual screenshots.

% \textbf{\sname mitigates the .}
% \textbf{Improve token efficiency.} Furthermore, \cref{tab:main_results} also shows that \sname substantially reduces average token consumption compared to the visual-memory baseline.
% Specifically, \sname uses {\color{blue} xx.x}\% fewer tokens per sample while still achieving better step accuracy on AgentNetBench.
% This efficiency is important for GUI agents because visual memory substantially increase context length, especially in large-screen GUI-cases.
% The result suggests that \sname provides a better accuracy-efficiency trade-off by preserving useful visual evidence while filtering out unnecessary screen information.

\textbf{Effectively leveraging visual memory.} Notably, simply adding visual memory does not always improve GUI agent performance.
For example, on OSWorld, the full-image memory baseline increases the hidden operation blindness rate from 67.1\% to 78.8\%, compared to the vanilla agent, indicating that storing full screenshots can introduce redundant or distracting context.
In contrast, \sname reduces it from 67.1\% to 60.3\%, outperforming both the vanilla agent and the full-image memory baseline.
This demonstrates that \sname does not merely add more visual history or information, but more effectively leverages visual memory by compacting and preserving task-relevant visual evidence.
% \sname also improves over the visual-memory baseline on AgentNetBench, achieving {\color{blue} xx.x}\% higher step accuracy while using {\color{blue} xx.x}\% fewer tokens.

% \textbf{Token Efficiency.}\input{resources/tables/4_1_token_efficiency}
% As shown in Table~\ref{tab:token_efficiency},  \sname substantially reduces average token consumption compared to the visual-memory baseline.
% Specifically, \sname uses {\color{blue} xx.x}\% fewer tokens per sample while still achieving better step accuracy on AgentNetBench.
% This efficiency is important for GUI agents because visual memory substantially increase context length, especially in large-screen GUI-cases.
% The result suggests that \sname provides a better accuracy-efficiency trade-off by preserving useful visual evidence while filtering out unnecessary screen information.

% \subsection{Component Analysis}
% \input{resources/tables/component_analysis}
% %\input{resources/tables/4_2_component_analysis}
% In this section, we provide a detailed analysis of \sname to validate the effect of each proposed component.

% \placeholder

% \subsection{Additional Analysis}
% 1. Independently compare retrieval, 2.  Max Step

% In this section, we provide additional analysis to show the specific characteristic of \sname.
% \placeholder

% \placeholder

\section{Case studies}
\label{sec:case_studies}
\Cref{figure:failure_concept} shows a representative \textit{hidden operation blindness} case in LibreOffice Impress. The task is to apply the \textit{Dissolve} transition to the first slide. Although the first slide is already selected, the transition settings are not directly visible in the main canvas and must be opened through a right-sidebar tab. The vanilla agent understands the high-level goal but guesses an incorrect route, clicking the top \textit{Slide} menu and a presumed transition submenu. Full-image memory provides a related prior state, but the useful cue is still embedded in a cluttered whole-screen observation, where the relevant sidebar icon is small and competes with many unrelated UI elements. 
As a result, full-image memory can suggest the correct route but does not reliably ground the action to the precise affordance. 
In contrast, the paired crop in action-grounded memory suppresses irrelevant visual context by isolating the sidebar tab or icon region. 
This changes the agent's next action from a menu-based guess to clicking the transition icon in the right sidebar in the current screenshot. 
Thus, the memory does not provide the final transition value directly, but supplies the hidden UI route. Through action-grounding, it makes the relevant visual affordance salient enough to produce the correct action.

Although action-grounded memory helps the agent interpret retrieved examples, it does not by itself guarantee recovery once the agent has entered an erroneous state. In an OSWorld VLC task, the agent must stream an HLS video by opening VLC's network-stream dialog and entering a \textit{.m3u8} URL. The vanilla agent follows the correct high-level route and reaches the target dialog but fails at the final low-level interaction by misclicking the \textit{Play} button and incorrectly assuming task completion. Action-grounded memory provides more interpretable UI regions and example actions, but it does not correct the faulty premise that the stream was closed without playback. AGMem resolves this error propagation through recovery-aware verification memory. When the detector flags the trajectory as erroneous, the recovery memory supplies examples of corrective behavior from similar faulty states, allowing the verifier to replace the unstable menu-click trajectory with the task-relevant recovery action \textit{Ctrl+N}, which directly reopens VLC's network-stream dialog, and succeeds the task. This case illustrates that action-grounded memory is useful for selecting and grounding normal actions, but recovery-aware memory is necessary when an earlier miss creates a misleading state premise that ordinary retrieval cannot correct.

% [TODO] Show concrete trajectory snippets for the two dominant failure causes identified by the per-image inspection on OSWorld:
%   (a) Image-state recognition failure: the agent misjudges which UI element is selected, whether a table is sorted, etc.
%   (b) Invisible operation: the operation requires a state not visible in the current screenshot (e.g., saving a file, opening a hidden menu).
% [TODO] Pair each failure case with its \sname recovery: how the action-grounded crop and/or the recovery memory steers the agent back on track.
% [TODO] Pick examples from OSWorld vanilla failures + AGMem successes for direct comparison.

\section{Related work}
\label{sec:related}

This section provides a comprehensive review of works related to \sname, including agentic memory, GUI agents, and compaction methods for multimodal agents.

\paragraph{Agentic memory.}
Recent works have studied how agents can reuse past experience during inference. Synapse stores past computer-control trajectories as textual examples and retrieves them for similar tasks~\citep{zheng2024synapse}. Agent Workflow Memory extracts reusable workflows from successful trajectories~\citep{wang2025agent}, and MemP builds procedural memory from step-level and script-level trajectory summaries~\citep{fang2026mempexploringagentprocedural}. A-Mem organizes memories as linked notes that can be updated over time~\citep{xu2026amem}.
Other works use failures as memory.
For example, ExpeL \citep{zhao2024expel} extracts text insights from success and failure pairs, and ReasoningBank~\citep{ouyang2026reasoningbank} stores reasoning strategies from both successful and failed runs.
These methods show that memory can improve agent decision-making. However, most of them store memory as text. For GUI tasks, text memory can miss important visual cues such as icons, layouts, selected regions, input fields, and error messages.
\sname instead stores action-relevant cropped images as memory. It also uses a separate recovery memory, so that the agent can retrieve corrective examples 
when it enters an erroneous state.

\paragraph{GUI agent.}
GUI agents aim to control graphical interfaces from screenshots and natural language instructions.
CogAgent~\citep{Hong_2024_CVPR} builds a VLM-based GUI agent with high- and low-resolution visual encoders.
% SeeClick~\citep{cheng2024seeclick} studies GUI grounding from screenshots and shows that accurate visual localization is important for GUI control.
OS-Atlas~\citep{wu2025osatlas} proposes a foundation action model with a unified action space, and Aguvis~\citep{xu2025aguvis} trains a pure-vision agent for autonomous GUI interaction.
OpenCUA~\citep{wang2026opencua} further provides open data, tools, benchmarks, and models for computer-use agents.
These works mainly improve GUI agents by training stronger models or collecting larger datasets.
% \sname is different from them because it does not update model weights.
% Instead, it improves test-time decision making by changing how past GUI experience is stored and retrieved.

Several GUI agents also use memory during planning.
\citet{agashe2025agent} proposed to use narrative memories for hierarchical planning.
% and RAP~\citep{kagaya2024rap} retrieves previous plans, actions, and observations as contextual memory.
CoMEM~\citep{wu2026autoscaling} compresses trajectories into continuous memory embeddings, and WebRAgent~\citep{zhang2026webragent} retrieves multimodal web trajectories for planning.
These methods are close to \sname because they also reuse previous GUI trajectories.
However, they usually use text summaries, full screenshots, trajectory-level memories, or learned embeddings.
% \sname stores each memory step as a subtask label, an executed action, and an action-relevant crop.
% This keeps visual evidence explicit while reducing unnecessary screen context.

\paragraph{Image processing.}
Visual focusing methods reduce irrelevant image regions for multimodal models.
Set-of-Mark~\citep{yang2023setofmark} adds visual marks to images to help grounding, and \citet{vstar} proposed to search images by zooming into useful regions.
CropVLM~\citep{carvalho2025cropvlm} learns a cropping policy for frozen VLMs.
For GUI agents, \citet{song2026compressfocusefficientcoordinate} proposed to reduce coordinate information during policy optimization, GUIPruner~\citep{xu2026spatiotemporaltokenpruningefficient} prunes visual tokens for efficient high-resolution GUI agents, and FocusUI~\citep{ouyang2026focusui} selects instruction-relevant visual tokens for GUI grounding.
While these works mainly focus on processing the current observation, \sname focuses on memories.
It extracts crop-view memories from past trajectory transitions and retrieves only the visual evidence that is related to the current subtask and state.
% Thus, \sname provides a memory-side way to make visual context compact and task-relevant.

\section{Conclusion and discussion}
\label{sec:con}

We took an analysis-first view of memory in GUI agents and showed
that screenshot-based agents can fail in benchmark-specific ways.
For example, with GPT-5.4-mini, OSWorld is dominated by \emph{cognitive failure} ($82.6\%$) and \emph{hidden operation blindness} ($67.1\%$), WebForge by \emph{grounding error} ($70.0\%$), and AgentNetBench by \emph{visual state misunderstanding} ($45.1\%$).
Adding na\"ive full-image visual memory does not move these dominant modes uniformly, while it helps where the bottleneck is recognition but amplifies template-style \emph{memory over-following} where the bottleneck is action-space reasoning.
\sname targets exactly these side effects: action-relevant crops keep memory references focused on the relevant UI region, and the recovery-state detector intervenes in residual cognitive failures, jointly yielding a $9.1\%$p OSWorld accuracy gain over the vanilla agent.
We see this analysis-design-reanalysis loop as a useful lens for future work on memory-augmented GUI agents.

\section*{Impact statement}

The main goal of this work is to advance the field of Machine Learning, in particular the analysis and design of memory for GUI agents.
As GUI agents observe whatever appears on a user's screen, deployments should filter sensitive UI content from any persistent memory and retain human oversight to prevent automated misuse.
Beyond these standard considerations, we do not feel any further societal consequences of our work should be particularly highlighted.

% Authors are \textbf{required} to include a statement of the potential broader
% impact of their work, including its ethical aspects and future societal
% consequences. This statement should be in an unnumbered section at the end of
% the paper (co-located with Acknowledgements -- the two may appear in either
% order, but both must be before References), and does not count toward the paper
% page limit. In many cases, where the ethical impacts and expected societal
% implications are those that are well established when advancing the field of
% Machine Learning, substantial discussion is not required, and a simple
% statement such as the following will suffice:

% ``This paper presents work whose goal is to advance the field of Machine
% Learning. There are many potential societal consequences of our work, none
% which we feel must be specifically highlighted here.''

% The above statement can be used verbatim in such cases, but we encourage
% authors to think about whether there is content which does warrant further
% discussion, as this statement will be apparent if the paper is later flagged
% for ethics review.

% In the unusual situation where you want a paper to appear in the
% references without citing it in the main text, use \nocite
\nocite{langley00}

\bibliographystyle{unsrtnat}
\bibliography{ref}
% \bibliographystyle{icml2026}

%%%%%%%%%%%%%%%%%%%%%%%%%%%%%%%%%%%%%%%%%%%%%%%%%%%%%%%%%%%%%%%%%%%%%%%%%%%%%%%
%%%%%%%%%%%%%%%%%%%%%%%%%%%%%%%%%%%%%%%%%%%%%%%%%%%%%%%%%%%%%%%%%%%%%%%%%%%%%%%
% APPENDIX
%%%%%%%%%%%%%%%%%%%%%%%%%%%%%%%%%%%%%%%%%%%%%%%%%%%%%%%%%%%%%%%%%%%%%%%%%%%%%%%
%%%%%%%%%%%%%%%%%%%%%%%%%%%%%%%%%%%%%%%%%%%%%%%%%%%%%%%%%%%%%%%%%%%%%%%%%%%%%%%
\newpage
\appendix
\onecolumn

\crefalias{section}{appsec}
\crefalias{subsection}{appsubsec}
\crefalias{subsubsection}{appsubsubsec}

\clearpage
\appendix

%%%%%%%%%%%%
\section{Preliminaries on memory-augmented GUI agents}\label{sec:appen-gui_intro}

A graphical user interface (GUI) agent is a large language model (LLM)-based autonomous agent that perceives and interacts with on-screen interface elements, such as screenshots, through human-like actions \cite{nguyen-etal-2025-gui}.

Given a long-horizon natural-language instruction $I$, the GUI agent operates over $T$ steps of observation-action pairs.
At each step $t \in \{0, 1, \dots, T-1\}$, the agent observes a full screenshot $o_t$ and executes an action $a_t$, after which the environment transitions to the next screenshot $o_{t+1}$.
From the initial screenshot $o_0$, we have the trajectory
$$ \tau := \bigl( o_0, a_0, o_1, a_1, \dots, o_{T-1}, a_{T-1}, o_T \bigr), $$
where $o_T$ is the final screenshot at which the goal of $I$ is achieved.

The agent can be formalized by an LLM-based policy $\pi$.
At each step $t$, the agent is given
(i) the instruction $I$,
(ii) a history
$h_t = \bigl((o_i,a_i)\bigr)_{i=\max(0,t-3)}^{t-1}$
of up to the three most recent screenshot-action pairs \cite{OSWorld},
(iii) the current screenshot $o_t$, and
(iv) a set of retrieved memory steps $\mathcal{R}_t$ from a memory bank $\mathcal{M}$, which stores previously executed trajectories.
The policy samples an action $a_{t}$, and the environment then transitions to the next screenshot $o_{t+1}$:
$$
    a_t \sim \pi\bigl( \cdot \big| I, h_t, o_t, \mathcal{R}_t \bigr),
    \quad
    o_{t+1} \sim P\bigl( \cdot \big| o_t, a_t \bigr),
$$
where $P$ denotes the unknown environment transition induced by executing $a_t$ on the device.
Because this policy operates over screenshots rather than HTML or accessibility trees, the framework is applicable to arbitrary desktop environments and applications \cite{OSWorld, cheng2024seeclick}.

%%%%%%%
\section{Algorithm details of \sname}
\label{app:algorithm_details}

In this section, we present our proposed framework \mname introduced in \Cref{sec:method_analysis} with details.
% , which addresses the two directions for improvement identified in Section~\ref{sec:main_results}.
% First, we formalize the problem setup and outline \sname (\cref{subsec:overview}).
% Then, we introduce salient memory entries that compact each memory entry to an action-relevant view (\Cref{subsec:cropping}).
% Finally, we describe salient memory candidates that narrow retrieval and provide a separate recovery memory for erroneous states (\Cref{subsec:recovery}).
Recall that an overview of \sname is depicted in \Cref{figure:method_overview}.

% In this section, we present our proposed framework \mname, a GUI agentic system that keeps GUI trajectory memory compact by retaining only salient parts.
% We first formalize the problem setup and briefly outline \sname in \cref{subsec:overview}.
% We then describe each component of \sname with details, in \cref{subsec:cropping,subsec:recovery}.
% \Cref{subsec:cropping} introduces \textit{salient memory entries}: each memory entry is reduced to an action-relevant view that retains only the screen region in which an action takes effect.
% \Cref{subsec:recovery} introduces \textit{salient memory candidates}: retrieval is narrowed to a small set of trajectories and steps aligned with the subtask structure, and a separate \textit{recovery memory} is used for recovery from erroneous states.
% An overview of \sname is depicted in \cref{figure:method_overview}.

% \input{resources/figures/4_method_overview}

% %%%%%%%%%%%
% \subsection{Overview}\label{subsec:appen-overview}

% \paragraph{\sname}

To guide the agent through a given long-horizon instruction $I$, \sname first decomposes $I$ into a list of \textit{subtasks} \cite{gao2024assistguitaskorienteddesktopgraphical,ye2025mobileagentv3fundamentalagentsgui}:
$$ \mathbf{s} = ( s_1, s_2, \dots, s_p ). $$
Here, each $s_i$ is a natural-language subtask derived from the instruction $I$ alone, before any action is executed.
Note that, the number of subtasks $p$ is generally different from the number of steps $T$, since one subtask may span multiple steps.
The pair $(I, \mathbf{s})$ is given to the agent throughout the trajectory.
Before emitting each action, the agent self-reports its current subtask, so that action execution is aligned with the current subtask.

\begin{figure*}[h!]
  \includegraphics[width=\textwidth]{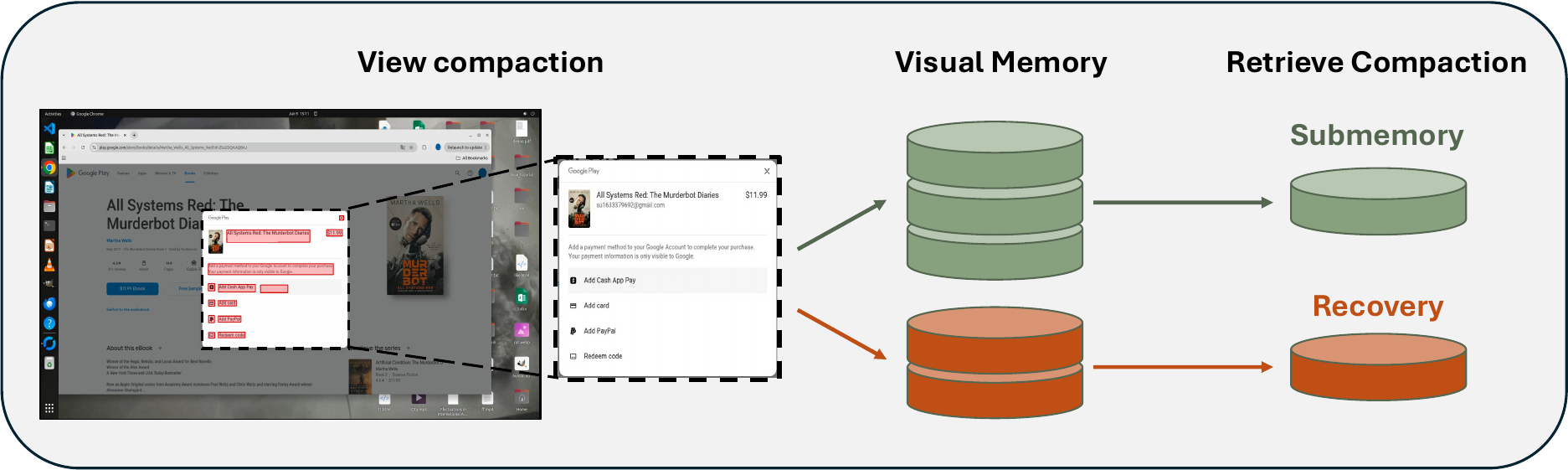}
  \caption{
  \textbf{Components of \sname.} \sname compacts GUI observations into structured memories, stores them in submemory and recovery modules, and retrieves compact context for future actions.
  }
  \label{figure:concept}
\end{figure*}

We then keep the memory salient in two core components.
First (\cref{subsec:appen-cropping}), every memory step is reduced to an \textit{action-relevant view}, a tight crop of the screenshot closely related to the executed action.
This action-relevant view is used for every step stored in memory, so that retrieval and action prediction can focus on action-relevant views rather than full screenshots.
Second (\cref{subsec:appen-recovery}), retrieval is restricted to salient candidates in two stages.
A trajectory-level retriever first selects a small pool of memory trajectories that cover the subtasks in $\mathbf{s}$.
Then, a step-level retriever searches for the few memory steps closest to the present state.
A recovery-state detector additionally redirects retrieval to a separate \textit{recovery memory} once the agent enters an erroneous state, providing examples of how to recover rather than only examples of successful execution.
\Cref{figure:concept} summarizes the main components of \sname.

%%%%%%%%
\subsection{Salient memory entries: action-relevant vision memory}\label{subsec:appen-cropping}

A desktop screenshot is often dominated by content unrelated to the current action or state (e.g., background windows, persistent toolbars, and idle panels), that are not affected by the agent's actions \cite{Lin_2025_CVPR, li2025screenspotpro}.
Storing all such screenshots would unnecessarily expand the memory bank and provide the agent's visual context with task-irrelevant cues.
We therefore reduce each memory entry to its salient form, which we call the \textit{action-relevant view}.
For each step $t$, we denote this view by $C_t$, which is a tightly cropped image patch that isolates the GUI region in which action $a_t$ takes effect.

\begin{figure*}[h!]
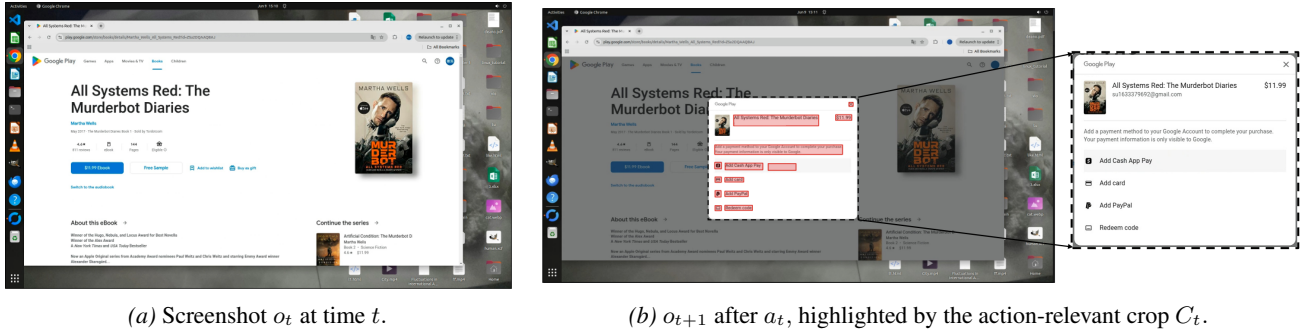

  \centering
  \begin{subfigure}[b]{0.39\textwidth}
    \centering
    \includegraphics[width=\textwidth]{images/before_crop.pdf}
    \caption{Screenshot $o_t$ at time $t.$}
    \label{fig:crop_input}
  \end{subfigure}
  \hfill 
  \begin{subfigure}[b]{0.59\textwidth}
    \centering
    \includegraphics[width=\textwidth]{images/after_crop.pdf}
    \caption{
    $o_{t+1}$ after $a_{t},$ highlighted by the action-relevant crop $C_t.$
    }
    \label{fig:crop_output}
  \end{subfigure}

  \caption{
    \textbf{Action-relevant vision memory.}
    Given a screenshot $o_{t}$ at time $t,$ \sname executes an action $a_{t},$ and then crop the action-relevant vision memory $C_{t}$ from $o_{t+1}.$
  }
  \label{figure:crop_overview}
\end{figure*}

\paragraph{Construction}
Given the screenshot $o_t$ (\cref{fig:crop_input}), the agent executes action $a_t$, and the environment transitions to the next screenshot $o_{t+1}$.
Based on the consecutive screenshots $(o_t,o_{t+1})$, we build the action-relevant view $C_t$ related to $a_t$ (\cref{fig:crop_output}).

% 프리앰블에 \usepackage{subcaption} 이 포함되어 있어야 합니다.

This construction proceeds in four steps.
First, we compute a pixel-wise difference between the two screenshots, subtract a locally smoothed version to suppress slowly varying background, and binarize the residual under a soft and a sharp threshold.
We apply standard morphological opening and closing to remove noise and connect fragmented regions into a single change mask~\citep{bhutada2022opening} .
Second, we take the tightest bounding box enclosing the change mask.
Third, we snap the bounding box to the smallest UI container, such as a window, panel, or dialog returned by a UI parser, that fully contains it.
Finally, if no qualifying container is returned by the parser, we keep the bounding box itself; if no change is detected at all, we fall back to the full screenshot.

At test time at step $t$, the action $a_t$ has not yet been executed, so $o_{t+1}$ and hence $C_t$ are not yet available.
We therefore use the previous step's view $C_{t-1}$, computed from the most recent pair $(o_{t-1}, o_t)$ by the same procedure, as the visual query for memory retrieval (see \cref{subsec:recovery}).
For $t = 0$, where no previous transition exists, we use the initial full screenshot $o_0$ directly as the query.

\paragraph{Memory structure}

We construct the memory bank $\mathcal{M}$ from AgentNet dataset \cite{wang2026opencua}, which we denote as $\mathcal{D}_{\text{AGNet}}$, where start from retaining only non-redundant and correct trajectories $\tau$. 

For each trajectory $\tau \in \mathcal{D}_{\text{AGNet}}$, an LLM produces a list of post-hoc subtask labels $\mathbf{s}_\tau = ( s_1, s_2, \dots, s_q )$ containing information about what actually happened on the screen during $\tau.$
In contrast to $\mathbf{s}$, which is derived from the instruction $I$ alone before any action is executed, $\mathbf{s}_\tau$ is generated post-hoc, after $\tau$ has been observed in full.
Hence, it is also grounded in the executed trajectory $\tau$ as well as the instruction $I.$
Thus, its segmentation reflects what actually happened on the screen through the trajectory $\tau$, rather than only what the user asked for through $I$.

We then segment $\tau$ along $\mathbf{s}_\tau$ into sub-trajectories.
Consequently, the atomic unit of memory is a \textit{memory step} defined as:
$$m_t = (s_j, a_t, C_t).$$
where $s_j\in\mathbf{s}_\tau$ is the subtask level that the step realizes, $a_t$ is the executed action, and $C_t$ is the action-relevant view.

\subsection{Recovery-aware memory retrieval}\label{subsec:appen-recovery}

Searching the entire memory bank at every step or action is both costly and noisy, since most trajectories are irrelevant to the current task and individual steps may produce spurious visual memories.
We therefore narrow the search space in two stages.
We first select a small pool of trajectories at the task level, and then retrieve individual memory steps within that pool.
In addition, when the agent flags an error, we redirect retrieval to a separate \textit{recovery memory}.

\paragraph{Two-stage retrieval}
Given the current subtask list $\mathbf{s}$, the trajectory-level retriever selects memory trajectories that cover the current task.
For each subtask $s \in \mathbf{s}$, we compare it with the subtask labels $\mathbf{s}_\tau$ of each memory trajectory $\tau \in \mathcal{M}$ using a Sentence-Transformer encoder \cite{reimers-2019-sentence-bert}.

A trajectory is considered relevant to $s,$ if any of its subtask labels is close to $s$ in the embedding space.
We greedily select a size-$k$ pool $\mathcal{M}_{\text{sub}} \subseteq \mathcal{M}$, where $k$ is the number of memory trajectories selected at the trajectory level.
This selection encourages the pool to cover all subtasks in $\mathbf{s}$ rather than repeatedly selecting trajectories for only a single dominant subtask.

After the trajectory pool is selected, the step-level retriever searches over the memory steps contained in $\mathcal{M}_{\text{sub}}$.
Each memory step is represented by both its subtask label and its action-relevant view.
We encode these with the CLIP encoders \cite{pmlr-v139-radford21a}: the subtask labels are encoded by the CLIP text encoder and the action-relevant view are encoded by the CLIP image encoder.
The two embeddings are combined with a fixed text-image weight of $0.5$ for each, which is shared across all retrieval queries.

The current state at step $t$ is represented in the same way, using the current subtask self-reported by the agent and the previous step's view $C_{t-1}$ as the visual query.
We then retrieve the top-5 memory steps whose combined representations have the highest cosine similarity to the current state, and append them to the agent's prompt as $\mathcal{R}_t$.

\paragraph{Recovery-aware verification}
One of the common failure modes of GUI agents is the error propagation.
Once the agent enters an erroneous state, its subsequent action proposals are made under a faulty premise.
This issue is difficult to handle with the main memory bank alone, because the main memory bank only stores successful executions and therefore does not contain examples of `how to recover' from mistakes.

To address this issue, we propose to use an LLM-based \textit{recovery-state detector}.
Before each proposed action is executed, the detector inspects the recent screenshots, the executed actions, and the proposed action.
It then predicts whether the agent is in an erroneous state and, if so, identifies the corresponding failure mode in \cref{tab:failure_modes}.
When no error is detected, the proposed action is executed as usual.
When an error is detected, the proposed action is blocked, and retrieval is redirected to a separate \textit{recovery memory}.
This recovery memory provides examples of corrective behaviors that bring the agent back from a faulty state to a valid one.

The recovery memory is constructed from AgentNet sub-trajectories that were excluded from the main memory bank as redundant or incorrect, but were later corrected within the same trajectory.
Each recovery example makes a pair of a faulty sub-trajectory and the corrective step that restores a valid state.
Each example is also labeled by failure mode, allowing retrieval to match the flagged failure pattern rather than rely only on global similarity.

%%%%%%%%%
\clearpage
\section{Experimental details}
\label{app:exp_details}

\subsection{Prompts for failure mode analysis}
\label{app:failure_prompt}

\begin{tcblisting}{
    colback=gray!5,
    colframe=gray!50,
    arc=5pt,
    boxrule=0.5pt,
    title=\textbf{Prompt},
    fonttitle=\bfseries,
    breakable,
    listing only,
    listing options={basicstyle=\ttfamily\small, breaklines=true}
}
You are given an evaluation result folder from a GUI/web/action-agent benchmark. Analyze the folder and produce failure-mode statistics using the taxonomy below.

The benchmark may be OSWorld, WebForge, AgentNet, or another benchmark with similar artifacts. Do not assume one fixed folder format. Inspect the folder first, infer the evaluation unit, and then apply the same root-cause definitions consistently.

## Input

The user will provide:

```text
RESULT_DIR=<absolute path to an evaluation result folder>
```

## Possible Artifacts

Use whatever files exist in the provided folder. Common artifacts include:

- summary files
- per-task or per-instance files
- trajectories
- raw model inputs/outputs
- screenshots/images
- memory/retrieval files
- ground-truth action files for offline benchmarks

Prefer raw model reasoning/output, executed actions, observations/screenshots, ground truth, retrieved memory, and final scores over pre-existing detector labels.

## Core Rules

Do not trust detector labels, evaluator names, superficial task types, or judge explanations as root causes by themselves. Use them only as pointers. The root-cause classification must come from the actual instruction, model reasoning, raw response, executed action, observation/image, ground truth, retrieved memory, and final outcome.

Use multi-label classification. One evaluation unit may have multiple root causes. Count each root cause at most once per unit.

## Unit Selection

Choose the evaluation unit based on the benchmark:

- Online task-completion benchmarks such as OSWorld and WebForge: use task-level units. One task is one evaluated instance.
- Offline action-prediction benchmarks such as AgentNet: use action-level units. One predicted action/step is one evaluated instance.
- If the benchmark is unknown, inspect the data:
  - If each record has a final task score or judge result, use task-level.
  - If each record compares a predicted action against a ground-truth action, use action-level.


## Denominator

Percentages must use the total number of evaluated units, not only failed units.

Use:

```text
ratio = count / total_evaluated_units * 100
```


Otherwise:

- Deduplicate repeated task IDs or action IDs when needed.
- Use the number of unique evaluated units found in the folder.

## Which Units To Label

Assign failure labels only to units that actually failed or partially failed.

Task-level benchmarks:

- Failed/partial means final score is below the success threshold, judge says incorrect, or result is false.
- If scores are numeric, use `score < 0.999` as the default failed/partial threshold unless the benchmark defines another threshold.

Action-level benchmarks:

- Failed means the predicted action is incorrect, mismatched, unparseable, or materially different from ground truth.
- If partial credit exists, treat non-perfect action scores as failed/partial unless the benchmark documentation indicates otherwise.

Successful units should not receive failure labels.

## Failure Categories

Classify each failed/partial unit into the following five categories.

### 1. Cognitive Failure

The model makes a wrong task-level, subtask-level, semantic, or progress-state decision.

Include cases where:

- It misunderstands the user's actual goal.
- It chooses the wrong next action or wrong subtask.
- It assumes a failed action succeeded and continues.
- It prematurely outputs `DONE`, `FAIL`, `WAIT`, or any equivalent terminal action.
- It keeps following an invalid plan despite evidence of failure.
- It makes a semantic/content reasoning mistake, such as wrong formula, wrong answer, wrong comparison, wrong extraction, wrong data transformation, wrong ordering, or wrong condition.
- It predicts an action that does not correspond to the ground-truth next action because it reasoned about the task incorrectly.

Examples:

- Ground truth is to click Search, but the model says it should click Settings because it misunderstood the intended next step.
- Task asks to calculate grades using a scale, but the model writes a formula with reversed thresholds.
- Task asks to approve a specific request, but the model opens or approves a different request.

### 2. Visual State Misunderstanding

The model misreads the current visible observation or screen state.

Include cases where:

- It thinks a dialog, menu, page, app, sheet, slide, layer, selection, cursor focus, checkbox, search result, object, or file is in a different state from what is visible.
- It says something is selected when it is not selected.
- It says a dialog is gone/open when the observation shows the opposite.
- It focuses on the wrong visible element because it misinterprets the screenshot/image.
- It predicts based on a nonexistent or misread visual state.

Do not classify as Visual State Misunderstanding merely because the unit failed. There must be evidence in the raw response/action or image that the model's perceived visible state differs from the actual visible state.

Examples:

- Raw response says "the image is selected", but the screenshot shows a text placeholder selected.
- The model describes a Settings page, but the observation is still a document editor.
- For an offline action step, the model targets a visible element that is not the one requested by the ground truth because it misread the image.

### 3. Hidden Operation Blindness

The model fails because the necessary operation is not directly visible in the current observation and must be found through a menu, submenu, dropdown, context menu, keyboard shortcut, command, file picker, save/export dialog, preferences/settings page, API call, or other hidden operation path.

Include cases where:

- The correct operation requires opening a menu or submenu.
- The correct operation requires a keyboard shortcut, command-line operation, or hidden app feature.
- The correct operation is behind Save As, Export, Import, Preferences, Settings, Format Cells, Chart Wizard, Extension Manager, file picker, or context menu.
- The model repeatedly tries visible UI controls but does not discover the hidden operation path.
- The model opens the wrong hidden path or cannot navigate a hidden menu/dialog workflow.

Do not classify as Hidden Operation Blindness just because the task is difficult. The failure must involve inability to find or execute a non-visible operation path.

Examples:

- Cannot find an export-as-PNG workflow.
- Cannot locate a setting buried inside preferences.
- Cannot use a required keyboard shortcut or command-line operation.

### 4. Grounding Error

The model's reasoning identifies the correct visible target, but the executed or predicted low-level action misses or fails to activate that target.

Only classify as Grounding Error when the raw reasoning shows that the intended target was correct.

Include cases where:

- The model says it will click the correct button/link/field, but the coordinates are off.
- It repeatedly clicks the correct intended target area and the observation does not change.
- It drags or selects the correct object/range in intent, but the actual selected region is wrong due to coordinate/drag grounding.
- Text entry targets the right field in reasoning, but focus is wrong due to click/coordinate failure.
- In action-prediction benchmarks, both prediction and ground truth intend the same element, but the predicted coordinate/bbox is outside the target.

Do not classify as Grounding Error when:

- The model chooses the wrong target from the start.
- The model reasons about the wrong element.
- The failure is due to not knowing the operation path.

Examples:

- Raw response says "click the close button", but repeated clicks at the coordinate do not close the dialog.
- Ground truth and prediction both refer to the Search button, but the predicted coordinate lands outside the button.

### 5. Memory Over-following

The model's current reasoning or action is contaminated by retrieved/example memory that does not belong to the current task context.

Only classify as Memory Over-following when there is explicit or highly concrete evidence that memory content influenced the wrong action.

Include cases where:

- Memory/retrieval files contain unrelated task details, and the raw response/action repeats those unrelated details.
- Raw response suddenly mentions a different application, file, object, target, keyword, color, number, menu, or workflow that appears in memory but not in the current task.
- The model follows an example-memory action sequence even though the current observation/task calls for a different action.

Do not classify as Memory Over-following when:

- Memory exists but the model does not follow it.
- The retrieved memory is relevant and helps solve the current task.
- The model merely selects the wrong subtask from the current task's own subtask list. That is Cognitive Failure, not Memory Over-following.
- There is no memory artifact in the folder.

Examples:

- Current task is in LibreOffice, but retrieved memory is VS Code auto-save and the model repeatedly opens "Visual Studio Code".
- Current task is to click a Search button, but the raw response mentions a remembered Settings workflow unrelated to the current observation.

## Classification Workflow

For each unique evaluation unit:

1. Determine whether the unit is successful, failed, or partially failed.
2. If successful, do not assign failure labels.
3. If failed/partial, inspect all available evidence:
   - instruction/task metadata
   - ground truth
   - raw model input/output
   - reasoning text
   - predicted/executed action
   - screenshots/images/observations
   - memory/retrieval content
   - final score/judge result
4. Assign all applicable root-cause labels from the five categories.
5. Count each category at most once per unit.

When deciding between categories:

- Wrong target because the model reasoned about the wrong UI element: Cognitive Failure or Visual State Misunderstanding.
- Correct target in reasoning but coordinate/bbox/action misses: Grounding Error.
- Failure to discover a menu/shortcut/dialog/workflow: Hidden Operation Blindness.
- Believing the current observation is in a state contradicted by the image: Visual State Misunderstanding.
- Wrong formula, wrong extraction, wrong answer, wrong ordering, wrong task decomposition: Cognitive Failure.
- Irrelevant retrieved memory leaks into current action: Memory Over-following.

## Required Output Format

First give one concise sentence describing:

- benchmark or inferred benchmark type
- unit type
- total evaluated units
- success count
- failed/partial count
- mean final score if available

Then output exactly this table format:

```markdown
| Root cause | Count | Ratio |
|---|---:|---:|
| Cognitive Failure | N | P% |
| Visual State Misunderstanding | N | P% |
| Hidden Operation Blindness | N | P% |
| Grounding Error | N | P% |
| Memory Over-following | N | P% |
```

Percentages must be computed as:

```text
count / total_evaluated_units * 100
```

Round percentages to one decimal place.

After the table, provide a short "Evidence Notes" section with 3 to 5 representative cases. For each representative case, include:

- category
- benchmark unit ID, task ID, or action ID
- raw response/action file path and line number if available
- screenshot/image filename if relevant
- memory file/path and line number if the case is Memory Over-following
- one sentence explaining why the classification is justified

Do not include long copied logs. Quote only the shortest raw content needed to justify the classification.

\end{tcblisting}
% You can have as much text here as you want. The main body must be at most $8$
% pages long. For the final version, one more page can be added. If you want, you
% can use an appendix like this one.

% The $\mathtt{\backslash onecolumn}$ command above can be kept in place if you
% prefer a one-column appendix, or can be removed if you prefer a two-column
% appendix.  Apart from this possible change, the style (font size, spacing,
% margins, page numbering, etc.) should be kept the same as the main body.
%%%%%%%%%%%%%%%%%%%%%%%%%%%%%%%%%%%%%%%%%%%%%%%%%%%%%%%%%%%%%%%%%%%%%%%%%%%%%%%
%%%%%%%%%%%%%%%%%%%%%%%%%%%%%%%%%%%%%%%%%%%%%%%%%%%%%%%%%%%%%%%%%%%%%%%%%%%%%%%

\end{document}